# Control of growth morphology of deposited *fcc* metals through tuning substrate-metal interactions

*Samuel Aldana*, Michael Nolan**

Tyndall National Institute, University College Cork, Lee Maltings, Dyke Parade, Cork T12 R5CP, Ireland

E-mail: samuel.delgado@tyndall.ie; michael.nolan@tyndall.ie

ABSTRACT

Precise control over thin film morphology is critical for optimizing material properties across diverse technological applications, as growth mode—whether 2D layer-by-layer or 3D island formation—determines key functional properties such as electrical conductivity in Complementary Metal-Oxide-Semiconductor (CMOS) interconnects and catalytic activity, where island distribution and size dictate performance. To explore the role of substrate interactions on metal morphology, we present extensive kinetic Monte Carlo simulations on six fcc metals growing in the (111) direction: Ag, Au, Cu, Ni, Pd and Pt. Our simulations enable screening and evaluation of growth modes under homoepitaxial scenarios and proposes morphology control strategies through substrate-metal interaction strength variation, modeled by modifying activation energies for atomic migration, combined with thermal vacuum annealing within typical back end of line (BEOL) integration thermal budget. Our results demonstrate that substrate interaction strength modulation effectively promotes either island formation or layer-by-layer growth, overcoming limitations in achieving large flat surface areas. Au, Pd and Pt exhibit the highest sensitivity to interaction strength variations, followed by Ag, showing that strongly interacting substrates decrease the root mean square (RMS) roughness, substrate exposure, island number and aspect ratios, with moderate increases in flat surface areas and atomic coordination numbers. Additionally, interconnect relevant metrics improve through thermal vacuum annealing particularly with strong metal-substrate interactions, reducing surface roughness, achieving larger flat areas, merging/smoothing islands, and decreasing defect density. We highlight not only the role of intrinsic metal migration barriers, but also the critical contribution of metal coordination number. Our results can support the selection of alloy components for target applications: Ni, Ag, Pd and Pt may be useful alloyed with other metals for interconnect applications, particularly given that Pd and Pt need to be combined with cheaper metals to leverage their beneficial properties while maintaining cost-effectiveness.

## INTRODUCTION

Controlling metal deposition at the nanoscale is essential to materials engineering and materials science, as it influences properties such as electrical conductivity, catalytic activity and optical response. The ability to modify the growth mode of metal deposition—whether two-dimensional (2D) or three-dimensional (3D)—is essential for a wide range of applications.[1] High-quality 2D metal films are critical for various devices, including photodetectors,[2] surface plasmon resonance biosensors,[3] and tunnel field-effect transistors.[4,5] However, surface defects such as roughness, 3D clusters, impurities, vacancies and dislocations can hinder charge transport and introduce scattering centers, increasing resistivity as the material is downscaled relative to its bulk properties.[6] While a 3D island growth mode degrades Complementary Metal-Oxide-Semiconductor (CMOS) interconnect performance, it can be used to promote catalytic activity, offering advantages for catalysis,[7] and playing a central role in the development of sustainable technologies and process intensification,[8] enabling more efficient energy conversion, chemical reactions and reduced footprint.

The continuous miniaturization of electronic devices demands not only high-quality crystalline metal deposition to maintain acceptable electronic properties, but also low temperature processing to preserve critical substrate characteristics. These include doping profiles, the interfacial integrity between layers and the prevention of metal diffusion into the substrate.[9] Achieving high-quality horizontal morphology of deposited metals at low temperatures remains a significant challenge in material science, as crystallinity and defect reduction typically require high-temperature annealing to facilitate atomic rearrangement of under-coordinated atoms into more stable configurations. However, excessive temperature can induce undesirable bulk diffusion or promote transition to island structures, ultimately compromising device performance. Consequently, precise control over metal deposition through metal-substrate interactions is key to ensure compatibility with CMOS process fabrication.

On the other hand, dealing with catalyst synthesis has different priorities, as the control over size distribution and the degree of dispersion of 3D clustering or nanoparticles is the key property that regulates catalytic activity.[10] Thus, while 3D clustering through sintering can be seen as beneficial for electronic devices applications since this promotes formation of a continuous layer, it is typically undesired in catalysis because it results in catalyst deactivation.[7,9] In catalysis, the preferred configuration is a dispersion of small 3D clusters or even single atoms,[11] although there are situations where the catalytic activity is size-dependent and the smallest clusters are not necessarily the best option.[12] It is important to note that some of the species employed in catalysis are precious metals, hence optimizing their use with respect to dispersion and size will have a significant impact in the cost of implementation.[13] Optimal use of catalysts means that the production of the desired products is maximized while the production of unwanted by-products is minimized.[14] Therefore, control of the distribution and dispersion of 3D clusters formed during the deposition of metals has a significant importance.[15]

The growth mode during metal deposition and annealing —whether 2D (horizontal) or 3D (island) — is strongly influenced by the metal-substrate interaction, which dominates atomic-scale kinetics.[16] Noble metals such as Ag, Au and Pt exhibit a natural tendency to form 3D clusters on weakly interacting substrates,[4,13] necessitating surface modification strategies to modify the morphology. One approach involves the use of gaseous species during the deposition, often called surfactants, to modify atomic diffusion and nucleation kinetics to enable the transition between 3D to 2D growth mode.[9] However, this method presents the risk of altering the physical properties of the noble metals layers—including electronic, optical, and transport properties, posing challenges for their use in metal-contact applications.[17] An alternative strategy is substrate engineering or the incorporation of additional liner materials to influence the atomic-scale kinetics. This strategy is particularly relevant for interconnect metallization in advanced CMOS device fabrication, where achieving high-quality 2D metal films is essential for minimizing resistance in the continuous miniaturization of integrated circuits.[18,19]

The control of metal deposition on 2D materials, such as graphene[4] and transition metal dichalcogenides (TMDs)[5] is critical for advancing next-generation of flexible electronics,[20] sensors,[21] optoelectronics,[22] and catalytic applications.[4] Their outstanding miniaturization potential and unique properties make them promising candidates for advanced device integration. Experimental studies have demonstrated the variation in metal growth mode on these substrates depending on the material combination. For instance, Ni, Au, and Ag exhibit 3D cluster formation on graphene, while Pt and Pd show similar behavior on graphene/Ru(0001). By contrast Au forms a continuous 2D monolayer on the same substrate.[4] In the case of $MoS_2$, Pd forms uniform contacts, Au arranges into isolated nanostructures, and Ag forms randomly distributed islands.[5]

In general, metal deposition on graphene and TMDs exhibits varying degrees of clustering depending on the relative strength of the metal-substrate interaction.[4,5] This interaction can be tuned controlling the number layers of the substrate[23] or by leveraging the graphene Moiré patterns formed on materials such as Ir(111)[24] or Ru(0001).[25] A major challenge in utilizing these materials for device applications is obtaining uniform 2D metal films without degrading their intrinsic electronic properties. This is particularly relevant for engineering Ohmic contacts in $MoS_2$-based transistors, where uniform metal coverage is essential for optimizing contact resistance and ensuring high-performance 2D electronic devices.[5]

In this work we examine how the as-deposited morphology of a series of *fcc* structured metals can be tuned, from 2D horizontal growth to 3D vertical island growth, through modulating the metal-substrate interaction in atomistic kinetic Monte Carlo deposition simulations. The substrate is not explicitly included in the simulation; instead, the substrate-metal interaction strength is modeled by modifying the activation energies for upward and downward migration of metal atoms. Scanning a series of activation/migration barriers that describe metal atom migration from substrate to metal and metal to substrate allows homoepitaxial growth and growth in conditions ranging from very weak metal-substrate interactions to very strong metal-substrate interactions to be simulated over realistic timescales (m*s* to μ*s*) at relevant processing temperatures. The selection of metals for this study (Au, Ag, Cu, Ni, Pd and Pt) is motivated by similar crystal structures (*fcc*), facilitating direct comparison within the same simulation model setup, and their technological relevance in applications such as catalysis, optoelectronics and CMOS interconnects.

For example, in contrast to bulk, dispersed Au nanoclusters exhibit unique catalytic properties,[26] whose electronic structure and cluster–substrate charge transfer depends significantly on the cluster size, bonding configuration and local environment, making the control of Au particles a relevant topic of study,[27] prompted by the discovery of low-temperature CO oxidation catalysis by supported Au nanoclusters.[28] Additionally, Au nanoparticles also play a significant role in photovoltaics[29] and biosensing.[30] Ag, valued for its excellent electrical, mechanical, and anticorrosive properties, is widely employed across diverse applications, including catalysis,[31] selective absorbers/emitters,[32] and high-temperature superconducting materials.[33] The ability to control Ag deposition is therefore of great interest,[34] particularly for optimizing performance in these fields.

Cu remains the primary material for interconnect fabrication in integrated circuits across all technology nodes due to its low resistivity and resistance to electromigration, enabling fast signal transmission and enhanced reliability. However, as interconnect dimensions continue to scale down, the formation of non-conducting 3D clusters poses significant reliability and integration challenges.[18] Cu morphology can be controlled by modifying its interaction with the underlying substrate with the aim to ultimately remove the extra seed layers used to promote Cu deposition.[19,35] While Cu and Ni are widely considered for CMOS interconnects,[36] alloys such as NiAl and $CuAl_2$ offer promising alternatives by eliminating the need for liner (seed) layers and diffusion barrier materials.[37,38] Understanding the morphological evolution of these materials is crucial for alloy selection, as their combination with other metals can yield superior performance. Additionally, some elemental metals from the platinum group, such as Pt and Pd, exhibit a less pronounced resistivity increase upon downscaling compared to Cu, making them potential candidates for certain interconnect applications.[36] However, their high cost remains a limiting factor in large-scale implementation. To address this, alloying strategies for Pt, Pd, and Ru, can leverage their advantageous electrical properties while reducing material cost by combining with less critical metals. This strategy is an option for performance optimization while maintaining economic feasibility.

First principles Density Functional Theory (DFT) calculations have proven effective in investigating various material properties, including the adhesion of metals to liner materials,[39] activation energies on different substrates,[18,40,41] and the thermal properties of metals.[42] Molecular Dynamics (MD), particularly *ab initio* MD (using DFT to evaluate energies and forces) is another widely used technique that can be used to determine activation energies and diffusion processes in elemental metals,[43,44] or metal alloys.[45] However, both DFT and MD are computationally expensive, and MD needs very long timescale simulations of at least tens of nanoseconds, outside of the reach of aiMD simulation methods, which limit their ability to realistically simulate large systems over macroscopic time scales. This constraint makes it challenging to simulate non-equilibrium processes such as film growth, which is dominated by infrequent atomic events and take place over longer time scales.

A cost-effective alternative is the use of mean-field approximations, which provide insights into long-timescale system evolution but lack atomic-scale resolution.[7] Given that the morphological evolution of film growth is determined by the relative kinetics of competing atomic-scale events, the kinetic Monte Carlo (kMC) algorithm is particularly well-suited for this purpose. KMC has been successfully employed to investigate the clustering of metals on weakly-interacting substrates,[16,35] 2D island morphologies,[46,47] and structural changes in nanoscale devices such as memristors.[48,49]

In this study, we employ kMC simulations to analyze the film morphology of six fcc metals growing in the (111) direction (Ag, Au, Cu, Ni, Pd and Pt) during homoepitaxial growth, the effects of varying substrate-metal interaction strengths and thermal vacuum annealing processes within typical thermal budget for BEOL integration. Activation energies for homoepitaxial growth are obtained from previous MD studies.[43,44] The combination of the substrate-metal interaction strength modulation and thermal vacuum annealing provides a promising strategy for tailoring film morphology according to target applications, particularly interconnect fabrication and catalysis. For interconnect applications, key metrics include surface layer root mean square (RMS) roughness, substrate exposure and flat surface areas. For catalytic applications, we analyze island coverage, the number of islands and island aspect ratios. Additionally, we examine the atom fractions with specific coordination numbers to estimate defect densities and occupation rate per layer, complementing previous metrics and providing a more comprehensive understanding of film morphology. Our simulation framework enables the analysis of film morphology across different metals under various realistic conditions, including diverse substrate interaction strengths and thermal vacuum annealing. This approach facilitates the development of morphology control strategies and provides a platform for screening and evaluating candidate materials for targeted applications such as catalysis and interconnect fabrication.

## METHODS

We employ the kMC algorithm to simulate the relative kinetics of competing atomic-scale events during film growth. The process includes atomic deposition and various types of migration, such as in-plane, upward and downward diffusion. These migration dynamics are influenced by the crystallographic facets the atoms move on and the nearest neighbors, as higher coordination numbers increase the activation energy. The algorithm consists of two main steps: (1) calculating the transition rates for all possible events in the current system state, and (2) a randomly generated number to select among the weighted events.

Atomic migration is a thermally activated process, with transition rates determined using transition state theory. This approach accounts for temperature dependence and the specific activation energy for the process, expressed as $\Gamma = \nu \cdot \exp(-E_A/K_BT)$,[50] where $\nu = 7 \times 10^{12}\ \mathrm{s}^{-1}$ is the pre-exponential factor and $E_A$ the activation energy of the corresponding event. However, atomic deposition is a non-activated process, with the transition rate determined by kinetic gas theory: $k_{ads} = P\sigma(T,\theta)A/\sqrt{2\pi m k_B T}$, where $P = 113\ Pa$ is the partial pressure of the gas, $T = 431\ K$ is the temperature, $\sigma$ is the sticking coefficient dependent on temperature and surface coverage ($\theta$), $m$ is the atomic mass, $A$ is the active surface area and $k_B$ is the Boltzmann constant.[51] The values employed for $P$ and $T$ are in line with those employed in Chemical Vapor Deposition for Cu.[52] According to the kinetic gas theory expression, increasing $P$ enhance the atomic deposition rate ($k_{ads}$), while increasing $T$ reduce it. However, while $P$ primarily affects the deposition rate without altering the relative probabilities of other events, T influences both the deposition rate and the relative probabilities of all competing processes. Specifically, elevated $T$ makes all events more likely to occur while reducing the disparity between low and high-probability events. Active area A is approximated by dividing the total simulation domain area by the number of adsorption sites. The sticking coefficient $\sigma$ is set to 1, independent of $T$ and $\theta$, a commonly used approximation.[51] Moreover, metals (e.g., Ag, Au and Cu) exhibit high sticking coefficients close to 1.[53]

The activation energies for atomic diffusion of Ag, Au, Cu, Ni, Pd and Pt during homoepitaxial growth on fcc (111) and (001) surfaces have been previously calculated employing MD.[43,44] The most relevant values are summarized in Table 1. To account for variations in activation energies arising from the local atomic environment, we express the activation barrier as: $E_A = E_k + E_{i,f}$, where $E_k$ is the kinetic barrier and $E_{i,f}$ $(\leq 0)$ is the energy difference between the initial and final sites, influenced by the number (and type in heteroepitaxy) of neighboring atoms. Transitions that increase the coordination number (CN) are energetically favorable, in which case $E_{i,f} = 0$, and the only barrier to overcome is the kinetic barrier. Conversely, transitions to less energetically stable lower-coordinated sites incur an additional energy cost. To model this, we apply a bond-counting scheme: $E_{i,f} = \max\left[(CN_f - CN_i) \times E_{CN}, 0\right]$, where $CN_{f,i}$ are the coordination numbers at the final and initial sites and $E_{CN}$ is the energy penalty per broken bond for each atomic species (see Table 1). For example, planar diffusion on a defect-free fcc(111) surface typically maintains a CN of 3 ($CN_f = CN_i$), resulting in $E_{i,f} = 0$. In contrast, a step ascent begins with a CN of 5 (3 from the lower layer and 2 in-plane), and the atom must detach from the 3 lower atoms, resulting in only two supporting the migration. This leads to a significant energy cost due to the CN reduction—for instance, the step ascent of Ag has a CN penalty of 0.645 eV, nearly the same value as the kinetic barrier of 0.62 eV.

To simulate the heteroepitaxial growth of different atomic species on various substrates, we consider a generic substrate that modifies the activation energies used in the homoepitaxial case (see Table 1). The substrate itself is not explicitly included in the simulation; instead, its influence is incorporated through adjustments to the migration barriers of the depositing atoms. We implement two approaches to capture the effect of substrate interaction strength: 1) the substrate affects step ascent and step descent in opposite ways, facilitating one and inhibiting the other; 2) the substrate affects step ascent migrations. All other activation energies remain unchanged. These scenarios allow us to isolate and evaluate the influence of step-related transitions on film morphology. Direct comparison across metal species is challenging because each metal species has a unique set of activation energies (see Table 1). To address this, we scale the same set of activation energies by a factor relative to the homoepitaxial case (increasing or decreasing depending on the case), ranging from 10% to 150% of the homoepitaxial activation energy. For example, in the case of Pd on a strongly interacting substrate (150% scaling) using the first approach, the step ascent barrier on the (111) facet increases from 0.068 eV (homoepitaxial) to 0.102 eV, while the step descent barrier decreases from 0.295 eV to 0.0295 eV. Under the first approach, only the step ascent barrier is modified. It is also important to note that in heteroepitaxial growth, the contribution of the substrate differs from that in the homoepitaxial case, where the ascending atom is bonded to three atoms of the same species in the lower layer. To model this substrate interaction for a generic substrate, we apply the same scaling factor to the homoepitaxial case. For example, in the Pd case, the substrate interaction ranges from -0.078 eV (10%) to -1.17 eV (150%). This contribution is included in the calculation of $E_{i,f}$ explained previously.

The simulation of these depositions on different substrates encounters a well-known limitation of the standard kMC algorithm: when the system becomes trapped in configurations dominated by low-energy barrier events isolated from the rest of the phase space by relatively high barriers—commonly referred to as superbasins. This situation frequently arises in cases such as intra-island diffusion on metallic surfaces, where the system undergoes several unproductive transitions within the superbasin, significantly decreasing the accessible simulation timescale.[54,55] A common but approximate solution is to artificially raise the lowest barriers; however, if it is not reasonably well equilibrated, this can corrupt the system dynamics.[54] To overcome this bottleneck, we adopt a more rigorous approach based on absorbing Markov chains. Although computationally more demanding, this method is exact and introduces no additional approximations. Local superbasins are identified on-the-fly by classifying states involved in low-barrier transitions as transient states, and those that lead to meaningful system evolution as bordering absorbing states.[55] To bypass the rapid, repetitive transitions within the superbasin, we construct a Markov transition matrix describing the probabilities among all transient and absorbing states. This enables the analytical calculation of both the exit probabilities and mean escape time from the superbasin, ensuring an accurate and efficient representation of the long-term kinetics. The approach accelerates the simulation if the analytical treatment is faster than waiting for an escape event through standard kMC. Finally, once the transition rates for exiting the superbasins and for all possible events are computed, the events are sorted using a binary tree structure, and the selected event is determined via binary search.[54,56] A second random number is then employed to calculate the time step, which is weighted by the total transition rate: $t = -ln(rand)/\sum \Gamma$, where *rand* is a uniformly distributed random number in the interval (0,1) and $\sum \Gamma$ is the summation of the transition rates for all available events.

**Table 1:** Activation energies (eV) for selected diffusion process of Ag, Au, Cu, Ni, Pd and Pt during homoepitaxial growth on 111 and 001 surfaces.[43,44]

| | Ag | Au | Cu | Ni | Pd | Pt |
|---|---|---|---|---|---|---|
| Terrace (111) | 0.064 | 0.117 | 0.043 | 0.061 | 0.109 | 0.171 |
| Terrace (001) | 0.467 | 0.531 | 0.477 | 0.376 | 0.621 | 0.875 |
| Step ascent: (111) | 0.62 | 0.089 | 0.311 | 0.304 | 0.068 | 0.153 |
| Step descent: (111) | 0.181 | 0.244 | 0.095 | 0.001 | 0.295 | 0.408 |
| Along edge: (111) | 0.302 | 0.237 | 0.309 | 0.385 | 0.381 | 0.461 |
| Along edge: (001) | 0.258 | 0.352 | 0.245 | 0.158 | 0.364 | 0.536 |
| CN contribution per atom | -0.215 | -0.18 | -0.26 | -0.212 | -0.26 | -0.32 |
| Substrate contribution (same metal) | -0.645 | -0.54 | -0.78 | -0.636 | -0.78 | -0.96 |

To simulate atom-by-atom adsorption and diffusion via kMC, we employ a discrete lattice model in which all possible atomic sites are predefined. The simulation domain consists of a $5 \times 5 \times 5$ nm face-centered cubic (fcc) lattice oriented along the (111) direction, with periodic boundary conditions applied in the lateral directions. The simulation is stopped when it reaches a target thickness of 1 nm. Simulations are performed for six different metal species: Ag, Au, Cu, Ni, Pd, and Pt. The lattice structures are generated using the Python Materials Genomics (pymatgen) library[57,58] and crystallographic data are retrieved from Materials Project using its API,[59,60] using the corresponding material identifiers: Ag (mp-124), Au (mp-81), Cu (mp-30), Ni (mp-23), Pd (mp-2) and Pt (mp-126). In a fcc crystal with (111) orientation, each atom has twelve nearest neighbors: six in-plane, three in the upper layer and three in the lower layer. As the film evolves, (111)- and (001)-like features emerge, each associated with distinct migration barriers (see Table 1). To identify these features, we compute the Wulff shape of each material using pymatgen,[61] and compare the geometry of each atomic surface and edge involved in a migration event. The main processes modeled are atomic deposition and surface diffusion on either the substrate or the evolving films. For a migration event to be allowed, the migrating atom must be supported by either the substrate or at least two nearest neighbors. For adsorption events, the atom needs either the substrate or three nearest neighbors. These conditions, which impose no explicit geometric constraints, allow for the natural formation of diverse three-dimensional morphologies.

## RESULTS

To investigate the impact of substrate interaction on film morphology, we performed kinetic Monte Carlo simulations as described in the Methods section, using the activation energies of the homoepitaxial case (Table 1) as a reference. To characterize the resulting morphologies, we employed key metrics: growth time to achieve a target thickness, RMS roughness (Figure S1a), reflecting surface morphology; substrate exposure fraction, representing the area of the substrate not covered by the film; layer occupation rate, indicating the growth mode (vertical vs. layer-by-layer); island size, defined as the number of atoms per island; normalized maximum flat surface area, corresponding to the largest flat region observed across all layers normalized to the simulation domain; island coverage fraction (Figure S1b and Figure S2), the area fraction covered by islands; island aspect ratio (Figure S1c); and atom fraction with CN of 9 and 12, used as a proxy for defect density. A combination of these metrics facilitates an assessment of the growth mode—whether it proceeds layer-by-layer, yielding continuous and smooth films suitable for interconnect applications, or forms dispersed 3D clusters, which can be a desired configuration for catalysis.

### Homoepitaxial Growth of *fcc* Metals

Figure 1 compares the homoepitaxial growth of six fcc metals (Ag, Au, Cu, Ni, Pd and Pt) using relevant metrics to characterize film morphology, while Figure 2 shows the corresponding simulated surface layers for each metal. The simulated growth patterns exhibit good agreement with previous experimental observations. For instance, the simulated Ag island formation exhibit similarities with previous experimental studies,[62] and the simulated Au layer morphologies align well with experimental growth patterns,[63] where early-stage deposition results in island formation that subsequently expands to form relatively flat terraces. Figure 1a presents the temporal evolution of the mean film thickness, allowing a direct comparison of growth time under identical pressure (P) and temperature (T) conditions. The time required to reach an average thickness of 1 nm depends on the adsorption rate ($k_{ads}$, see Methods section) and growth mode, as metals favoring vertical island formation exhibit a more rapid increase in mean thickness. The corresponding adsorption rates are: $k_{ads-Ag} = 10.1 \cdot 10^6 s^{-1}$, $k_{ads-Au} = 7.72 \cdot 10^6 s^{-1}$, $k_{ads-Cu} = 9.97 \cdot 10^6 s^{-1}$, $k_{ads-Ni} = 9.77 \cdot 10^6 s^{-1}$, $k_{ads-Pd} = 9.39 \cdot 10^6 s^{-1}$ and $k_{ads-Pt} = 6.90 \cdot 10^6 s^{-1}$. Among the six metals, Pt exhibits the longest growth (0.75 μs) time, while Ag shows the shortest (0.54 μs), consistent with the higher $k_{ads}$ of Ag compared to Pt. Despite this difference in growth times, both show similar RMS roughness (3.45 and 3.41 Å, respectively). Cu, Ni and Pd show similar adsorption rates and RMS roughness values (3.7, 3.5 and 3.9 Å), resulting in comparable growth times (0.64, 0.67 and 0.59 $\mu s$). In contrast, although Au has a significantly lower $k_{ads}$, its larger RMS roughness (4.13 Å) compensates for the slower adsorption, yielding a growth time (0.63 $\mu s$) similar to that of Cu, Ni, and Pd.

It is important to note that low RMS values alone do not necessarily indicate a flat surface, which is the desired outcome for interconnects. A surface composed of small, uniformly distributed islands and valleys can also yield low RMS values, a morphology typically preferred in catalysis. This distinction is evident when comparing the surface morphologies of Au and Pt: although Au exhibits higher RMS roughness due to deeper valleys, their surface features differ significantly (Figure 2b vs. Figure 2f), with Au displaying a larger normalized maximum flat surface area (Figure 1f). Therefore, while RMS roughness is a helpful indicator of surface irregularity, it needs to be complemented with other metrics.

Figure 1c presents the atom fraction as a function of their CN, offering insight into the defect density of the film. In an ideal fcc crystal growing along the (111) direction, each atom has six in-plane neighbors, three in the layer above and three in the layer below, yielding a total CN of 12. Atoms in the first and last layers, lacking either the upper or lower coordination, exhibit a maximum CN of 9, excluding the substrate. A perfect 1-nm-thick film consists of five atomic layers: three composed of atoms with a CN of 12, and two with atoms exhibiting a CN of 9. Atoms with CN below these expected values indicate structural defects such as vacancies, while partially filled final layers suggest island formation. In contrast, an abrupt drop in layer occupation reflects the formation of smoother and continuous films. Figure 1d shows the atom fraction with a CN = 12 for each metal relative to the corresponding ideal crystal structure. Ni shows the highest atom fraction with CN = 12 (0.89), suggesting the formation of a high-quality film with fewer defects and a layer-by-layer growth mode, followed by Cu (0.84) and Pd (0.7). A low concentration of undercoordinated atoms is essential in interconnect fabrication, as they act as scattering centers that increase resistivity.[6] Conversely, undercoordinated atoms can be advantageous in catalysis due to their dangling bonds.[64] This consideration is relevant when selecting metals with a high fraction of undercoordinated atoms as alloying elements, such as Ag, Au, Pd and Pt (Figure 1d), as this may limit their suitability for interconnect applications. This limitation can be mitigated by employing strongly interacting substrates that suppress undercoordination or through post-deposition annealing, as demonstrated later in this study.

To identify islands formed during growth, the film is first divided into horizontal slices corresponding to each atomic layer. Within each layer, a slice is defined as a contiguous group of atoms connected through in-plane nearest neighbors. Islands are then constructed by linking slices from adjacent layers if they share at least one vertical nearest-neighbor connection. The reference is the base of the islands, i.e., the last continuous layer, above which discrete island formation occurs. It is determined as the highest layer (from bottom to top) that contains only a single slice. If all layers consist of a single slice, the reference layer is chosen as the highest one with an occupation rate below 80%. Once the islands are identified (e.g., for Pd in Figure S2), their mass is calculated by counting the number of atoms they contain (Figure 1e). Pt exhibits the largest island (393 atoms) among the metal studied, with additional smaller features present including an isolated atom and a small 3-atom cluster, see Figure 1e and 2d. Ag and Cu also form relatively large islands, as reported previously,[65,66] with average sizes of $173 \pm 20$ and $148 \pm 92$ atoms, respectively, compared to the smaller islands formed by Ni ($78 \pm 14$ atoms), Pd ($67 \pm 30$ atoms) and Au ($46 \pm 17$ atoms). Larger islands result in a lower surface-to-volume ratio and promote the coalescence of neighboring islands, favoring the formation of continuous films required for interconnects. Furthermore, Ostwald ripening can also contribute to reducing the overall surface-to-volume ratio, where smaller, less stable islands dissolve and their material is deposited onto larger, more stable islands. In contrast, the higher dispersion and greater surface-to-volume ratio of the smaller islands observed for Ni ($78 \pm 14$ atoms), Pd ($67 \pm 30$ atoms) and Au ($46 \pm 17$ atoms) are advantageous for catalytic applications.

The normalized maximum flat surface area (Figure 1f) is defined as the total area of a slice not covered by atoms from any upper slice. Au exhibits the largest maximum flat surface area, a feature generally favorable for interconnect applications. However, it also displays a low atom fraction with CN = 12 (Figure 1d), small island sizes (Figure 1e) and the highest substrate exposure fraction, three characteristics detrimental to interconnect applications. Cu presents a high proportion of atoms with CN = 12 (Figure 1d), relatively large islands (Figure 1e) and low substrate exposure fraction, although its maximum flat surface area is moderated. Ag and Ni also show large flat surface areas, although with lower atom fraction with CN = 12 and smaller island sizes compared to Cu. In the following sections, we explore how substrate-metal interaction strength and thermal vacuum annealing can be used to enhance these morphological features.

Figure 1g displays layer-by-layer occupation rate at the end of the simulation. In an ideal film, each layer would be fully occupied (100%), with an abrupt drop to zero beyond the final layer. In practice, partial layer occupation is observed, and sparsely populated upper layers indicate the presence of islands. Among the studied metals, Au exhibits the lowest occupation rate in the first two layers (~90%), corresponding to a high substrate exposure fraction of 0.09 (see Figure 1h). In contrast, Ag, Cu, Ni, Pd and Pt maintain higher occupation rates (> 94%) in the initial layers, with a progressive decline in subsequent layers. While Ag shows high initial populated layers, its occupation rate decreases more rapidly compared to the other metals. Cu and Ni have a slower decrease, with the highest occupation rates in the first two layers. This is reflected in their substrate exposure fraction (0.0009 for Ni and 0.008 for Cu), the lowest among all metals studied (Figure 1h). Overall, the metals with the lowest substrate exposure fraction—Ni, Cu, and Ag—align with those currently used in interconnect technologies.

The observed differences in film morphology (see the key metrics in Figure 1 and the surface layers in Figure 2) among the metals studied can be directly related to differences in their atomic migration barriers. A comparison between Au and Pd illustrates this: both exhibit similar activation energies for step ascent (0.089 eV for Au, 0.068 eV for Pd) and descent (0.244 eV for Au, 0.295 eV for Pd), suggesting a comparable preference for upward migration. However, Pd has a significantly stronger CN contribution to the activation energy ($-0.26\ \mathrm{eV/atom}$) than Au ($-018\ \mathrm{eV/atom}$), indicating that atomic detachment is more energetically unfavorable for Pd. Consequently, Pd atoms find it more difficult to reduce their CN compared to Au, which explains Pd's superior film quality—lower RMS roughness, reduced defect density, lower substrate exposure, and larger island mass.

When comparing Pt and Au, both show low step ascent barriers (0.153 eV for Pt, 0.089 eV for Au), but Pt exhibits the highest step descent barrier among the studied metals (0.408 eV), making upward migration favourable and downward migration highly unfavorable. Additionally, Pt shows the strongest CN contribution ($-0.32\ \mathrm{eV/atom}$), which hinders atomic migrations that would reduce CN. These factors result in Pt exhibiting the highest overall film quality among Au, Pd and Pt, characterized by the lowest RMS roughness, low substrate exposure fraction, the largest island size, and a relatively large maximum flat surface area, while maintaining a fraction of atoms with CN = 12 similar to Ag. This comparison highlights the critical role of CN contribution on the resulting film morphology.

Ag and Ni show similar CN contribution ($-0.21\ \mathrm{eV/atom}$) but differences in step ascent and step descent energy barriers. Ag has the highest step ascent barrier (0.62 eV) and a moderate step descent barrier (0.181 eV), while Ni shows a negligible descent barrier Ni (0.001 eV). As a result, Ni exhibits some superior morphological characteristics, specifically in the reduced defect density (highest fraction of atoms with CN = 12), higher occupation rates in the initial layers and the lowest substrate exposure fraction, while maintaining comparably low surface roughness. This highlights the relevance of the step descent barrier in cases with equivalent CN contribution.

Finally, although Cu has a lower step descent barrier than Ag (0.095 eV), but higher than Ni, it exhibits a stronger CN contribution (-0.26 eV/atom) compared to both Ni and Ag. This results in Cu achieving the second highest fraction of atoms with CN = 12, large island sizes, the highest occupation rates in the initial layers and the second lowest substrate exposure. Nonetheless, Ni still outperforms Cu in terms of atom fraction with CN = 12, substrate coverage, with a comparable high occupation rate for the initial layers. Overall, these comparisons reveal the complex interplay between the different atomic migration energies—step ascent, step descent, and CN dependence—and their impact on film morphology. This complexity underscores the necessity of long-time scale kinetic simulations to predict growth behavior, as simple extrapolations from individual migration energies are insufficient.

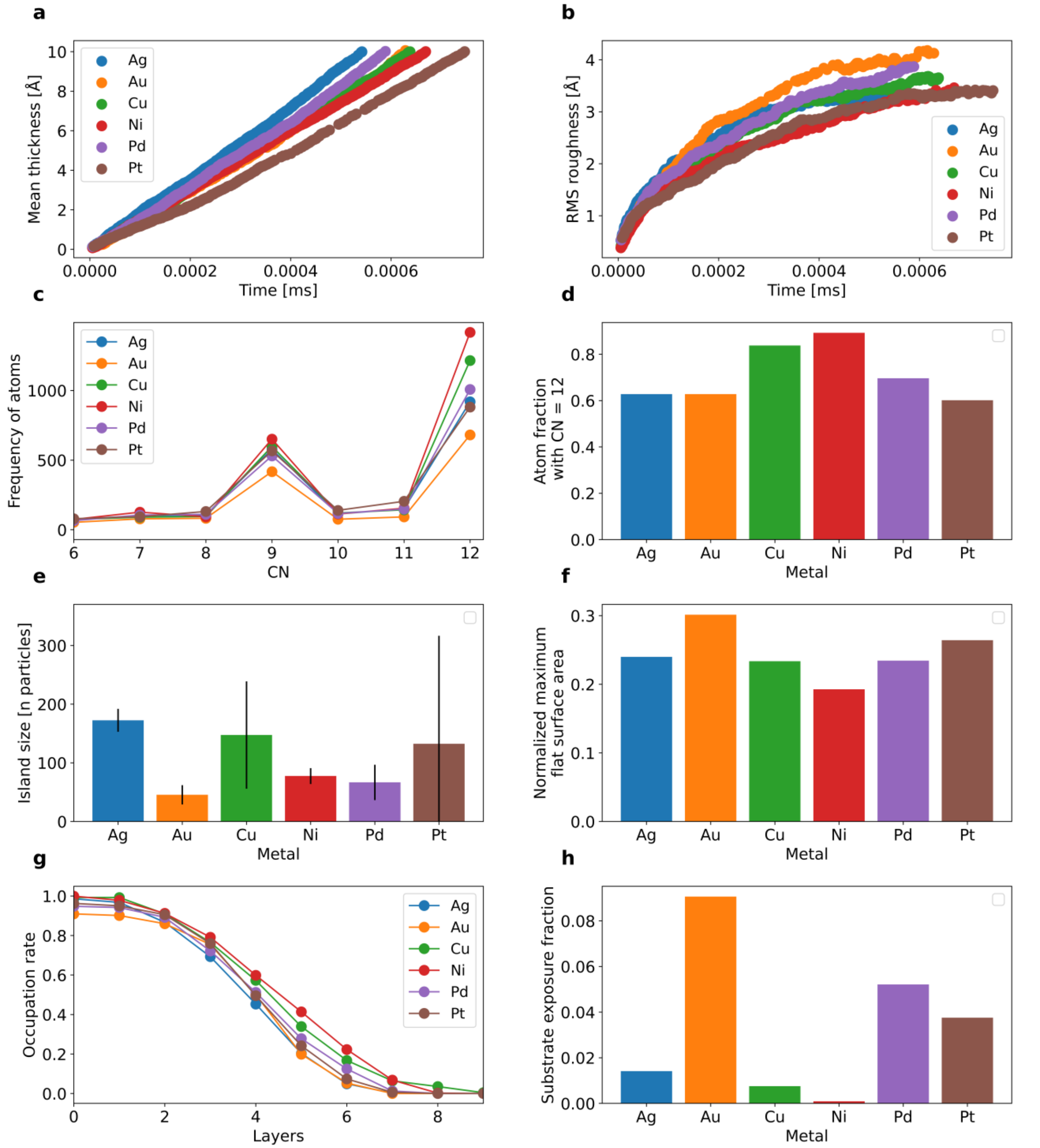

**Figure 1.** Comparison of homoepitaxial growth of six fcc metals: Ag (blue), Au (orange), Cu (green), Ni (red), Pd (purple) and Pt (brown). a) Temporal evolution of the mean thickness; b) temporal evolution of the RMS roughness; c) frequency of atoms as a function of CN; d) atom fraction with 12 nearest neighbors e) mean island mass with the black line the standard deviation, f) normalized maximum flat surface area; g) occupation rate per layer; and h) substrate exposure fraction.

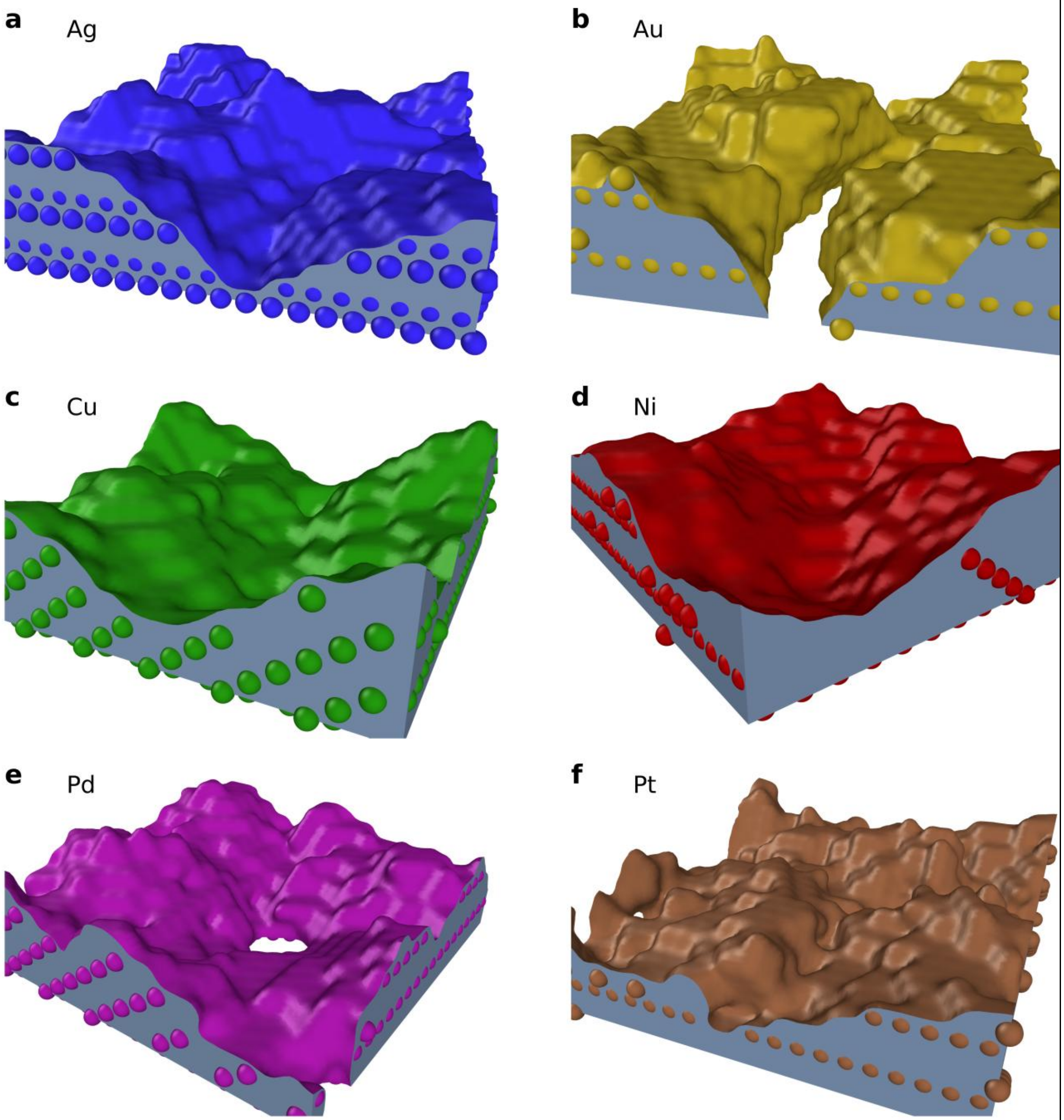

**Figure 2.** Surface layer resulting from homoepitaxial growth of (a) Ag, (b) Au, (c) Cu, (d) Ni, (e) Pd, and (f) Pt.

**Impact of Substrate-Metal Interaction Strength on Metal Morphology**

We investigate the influence of substrate interaction strength by modifying the step ascent and descent barriers relative to the homoepitaxial case, modeling the initial kinetic effects arising from the thermodynamic minimization of the system's Gibbs free energy. Weak substrate interactions energetically favor configurations where metal atoms are less bound to the substrate, which effectively lowers the step ascent barrier and raises the step descent barrier, hence promoting upward migration and island formation. Conversely, strong interactions energetically favor configurations where metal atoms are strongly bound to the substrate, effectively increasing the step ascent barrier and decreasing the step descent barrier, favoring downward migration and promoting 2D growth. While this approach sets the initial kinetics based on the substrate interaction, it should be noted that as the film grows thicker, the effective barriers may evolve towards the homoepitaxial values. However, for the initial growth stages simulated here (e.g., up to 1 nm thickness), we assume that the substrate's influence remains dominant and the constant barrier approach is a valid approximation. A compilation of activation energies for each material and each substrate-metal interaction strength is presented in Figure 5, with step ascent barriers in the x-axis and step descent barriers in the y-axis. Notably, Pd, Au and Pt have the lowest step ascent barriers: 0.068, 0.089 and 0.153 eV, respectively. Even at the strongest interaction strength (150% of the homoepitaxial metal-substrate interaction), Pt's barrier (0.229 eV) remains comparable to those of Ni and Cu at 70% (~0.213 eV). At 50% of the homoepitaxial interaction strength, Pt's barrier (0.077 eV) is similar to Pd's homoepitaxial case (0.068 eV). In contrast, Ag, Cu and Ni have the lowest step descent barriers (0.181, 0.095 and 0.001 eV, respectively). At the weakest substrate interaction, Ag's step descent barrier becomes comparable to those of Pd and Au in the homoepitaxial case and significantly lower than that of Pt. For reference, Au at the weakest interaction strength exhibits a step descent barrier similar to that of Pt in the homoepitaxial case. The corresponding surface layer morphologies across the substrate-metal interaction strength range are shown in the Supporting Information: Ag (Figure S3), Au (Figure S4), Cu (Figure S5), Ni (Figure S6), Pd (Figure S7) and Pt (Figure S8).

Figure 3 shows how key metrics for the deposited metals that are relevant to interconnect—RMS roughness, the substrate exposure fraction (with respect to the total area) and the normalized maximum flat surface area (respect to the total area)— depend on surface interaction strength. Figure 4 focuses on metrics more pertinent to catalysis, such as the total area covered by the islands (indicative of catalytically active region), number of islands (distinguishing whether coverage results from a few large clusters or from numerous small ones, where a high dispersion of small islands is generally preferred for catalysis), and aspect ratio (height-to-width, reflecting island geometry, with higher values indicating sharper islands and lower values indicating flatter morphologies). Some data points for the weakest metal-substrate interactions cannot be included due to the high computational cost of simulations with the low energy barrier problem, described in the Methods, which requires absorbing Markov chains. In some cases—such as Au—the substrate interaction is too weak to even retain the metal on the surface.

Figure 3a shows that increasing the substrate-metal interaction strength from 10% toward the homoepitaxial value leads to a significant reduction in surface roughness for Pt, Pd and Au—from initial values close to 9.1, 13.5 and 10.7 Å, respectively, down to ~2.5–3.5 Å. Ag also exhibits a slight reduction in roughness, from 4.2 Å to a range of 2.8–3.6 Å, indicating a weaker dependence on interaction strength. These values converge toward the consistently low roughness observed for Ni, Cu and Ag across all interaction strengths. The reduced roughness in these metals is likely associated with their relatively low step descent barriers, ranging from 0.181 eV (Ag) to 0.001 eV (Ni), consistent with previous studies showing that enabling step descent migration promotes smoother film growth.[18,67] A similar trend is observed for the substrate exposure fraction (Figure 3b): values for Pt, Pd and Au decrease markedly from 0.3–0.76 on weakly interacting substrates (≤50%) to 0.005–0.02 at 90% interaction strength, close to the low values observed for Ni, Cu and Ag across all interaction strengths. Ag again exhibits a weaker dependence on interaction strength, with its substrate exposure fraction decreasing from 0.07 at 30% to 0.01 at ≥50% interaction strengths.

The impact of substrate interaction on both RMS roughness and substrate exposure is strongest for Au, followed by Pd and least pronounced for Pt. In contrast, Ag, Cu, and Ni show minimal to negligible sensitivity. This trend may be attributed to their respective CN contributions to activation energy (-0.18, -0.212 and -0.32 eV/atom), combined with their relatively low step ascent and high step descent barriers, which promote upwards migration. For example, at 10% interaction strength, Pt exhibits RMS roughness of 9.1 Å and substrate exposure of 0.3, lower than Pd (13.5 Å and 0.67) and Au (10.7 Å and 0.76) at 50% interaction strength. In fact, at 50% interaction strength, Pt already shows markedly lower values of 3.7 Å and 0.05, underscoring its reduced sensitivity to the substrate interaction strength.

Trends in the normalized maximum flat surface area (relative to the total area) can also serve as an indicator of island formation. Islanding is evident for Au (Figure S4a), Pd (Figure S7a–c) and Pt (Figure S8a–b), corresponding to low normalized maximum flat surface areas (≤0.2) on weakly interacting substrates, as shown in Figure 3c. In contrast, Ag, Cu and Ni do not exhibit a clear dependence on interaction strength, with values ranging from 0.22 to 0.31 across the entire range. As seen in Figure 3c, none of the metals achieve flat surface areas exceeding one-third of the simulation domain after deposition—well below the ideal case. However, this morphological feature can be significantly improved through thermal vacuum annealing, as we will discuss later.

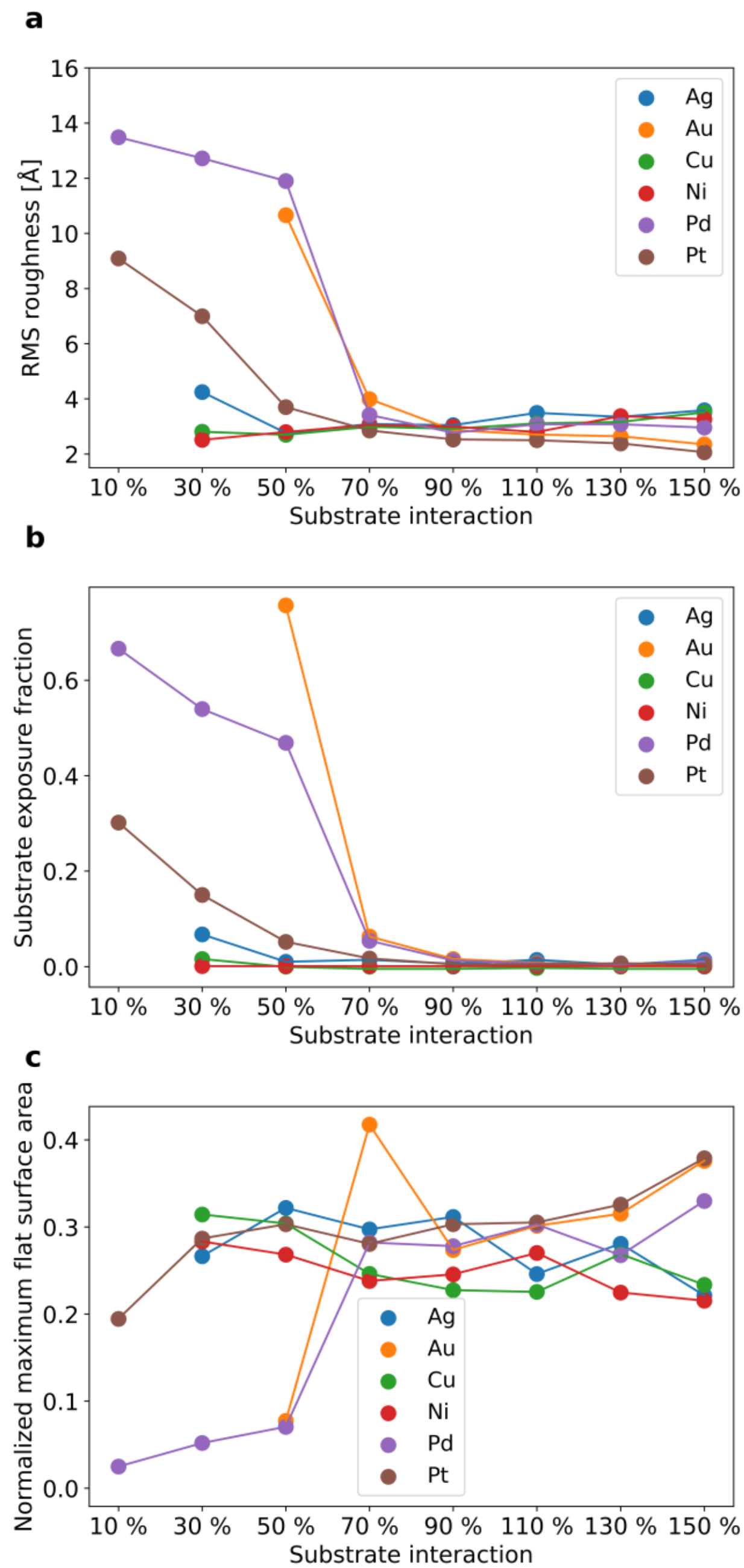

**Figure 3.** Influence of substrate-metal interaction strength (ranging from 10% to 150% of the homoepitaxial step ascent and descent barrier) on surface morphology metrics relevant to interconnects for Ag, Au, Cu, Ni, Pd and Pt: (a) RMS roughness, (b) substrate exposure fraction relative to the total area and (c) normalized maximum flat surface area relative to the total area.

Figure 4 presents morphology metrics relevant to catalysis: the island coverage fraction (Figure 4a), number of islands (Figure 4b) and mean aspect ratio (Figure 4c). To maintain the clarity of the main plot in Figure 4c, the standard deviation of the aspect ratio is presented separately as an inset. These metrics should be interpreted together to evaluate the morphology of the metals and how this can relate to catalytic activity. An ideal situation would combine a high total area covered by many small, sharp islands—reflected in high aspect ratio values. Among the studied metals, there is no clear trend for island coverage fractions across substrate interaction strengths. Notably, within this interaction strength range, Ag, Cu and Ni form fewer than three islands with average aspect ratios below 0.32, indicating that a small number of flat islands dominate the surface morphology, which would not be favourable for catalysis. In contrast, Au, Pd and Pt present characteristics more suitable for catalysis applications: a higher number of islands with relatively large aspect ratios on weakly interacting substrates. For instance, Pd shows the most promising features for catalytic applications, forming 25–27 islands with average aspect ratios of 1.57-1.60 at 10–30% interaction strength, followed by Au with 18 islands and an average aspect ratio of 1.39 at 50%. Pt forms 18 and 7 islands with average aspect ratios of 0.70 and 0.51 at 10% and 30% interaction strength, respectively. Figure 4b and 4c show a clear decrease in both the number of islands and mean aspect ratio with increasing substrate-metal interaction strength, reaching values characteristic of Cu, Ni and Ag for interaction strength ≥70%—namely, 1-3 islands and aspect ratios around 0.3. This behavior, similar to the trends observed in RMS roughness (Figure 3a) and substrate exposure fraction (Figure 3b), indicate that stronger interactions—associated with higher step ascent and lower step descent barriers—inhibit islanding and promote flatter film morphologies.[18,35,68] This trend is less pronounced in Cu, Ag and Ni, likely due to their intrinsically high upward and low downward migration barriers.

In particular, Ni consistently forms only 1-3 islands across all interaction strengths (Figure 4b) and displays one of the lowest mean aspect ratios (Figure 4c), indicative of broad, flat islands. This morphology is unfavorable for catalytic applications, as previous studies have shown that extended Ni surfaces are catalytically inactive, with only low-loadings exhibiting activity.[69] For Pd, substrate-metal interaction strengths between 10% and 50% is the best range for achieving island distributions, while higher values (>50%) favor the formation of flatter surfaces. In contrast, Pt requires a narrower interaction strength range (10%-30%) to generate small island distributions. Au exhibits the most restricted range (50%-70%), as interaction strengths below this threshold are insufficient to maintain material adhesion to the substrate, while stronger interactions (>70%) promote flattening and result in surfaces with low aspect ratios and fewer islands. Ag also exhibits a moderate number of islands with slightly larger aspect ratio (although smaller than those of Au, Pd and Pt) within a substrate interaction strength range of 30%-50%. However, more pronounced effects would likely be observed at interaction strengths below 30%, a trend that also applied to Cu and Ni, which show minimal morphological changes even at the lowest simulated values. Unfortunately, simulating these lower interaction strength conditions for these metals presents significant computational challenges due to the associated high computational cost due to persistent low barrier problems.

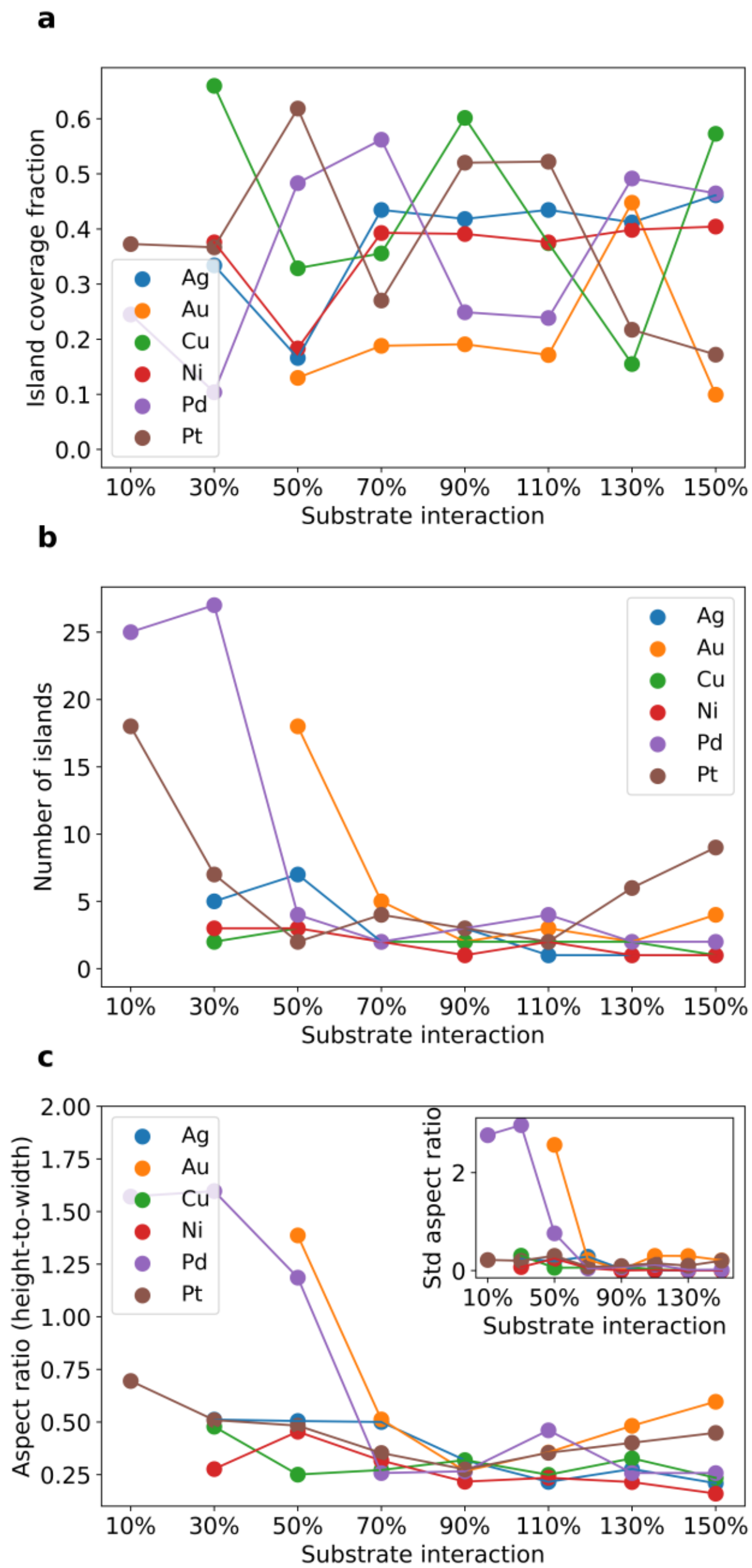

**Figure 4.** Influence of substrate-metal interaction strength (ranging from 10% to 150% of the homoepitaxial step ascent barrier) on surface morphology metrics relevant for catalysis for Ag, Au, Cu, Ni, Pd and Pt: (a) island coverage fraction, (b) number of islands and (c) mean aspect ratio of islands (height-to-width). Inset: the standard deviation of the aspect ratios.

To provide a comprehensive global perspective of the transition from island formation to flattened surfaces, we developed a quantitative metric defined as the product of the island number and the average aspect ratio. High values indicate either numerous small islands (high island density with low individual aspect ratios) or fewer elongated islands (moderate island density with higher individual aspect ratios), both of which represent favorable configurations for catalysis. Conversely, low metric values correspond to few, broad, flat islands, which are generally unfavorable for catalytic applications. Figure 5 illustrates this transition using a heatmap representation of the metric (island number × average aspect ratio), with the step descent barrier plotted on the y-axis and the step ascent barrier on the x-axis, both of which are directly related with substrate interaction strength. Consistent axis limits and color bar scales have been maintained across all materials to enable direct comparison. As previously discussed, Ag (Figure 5a), Au (Figure 5b), Pd (Figure 5e) and Pt (Figure 5f) demonstrate that weak substrate interactions (≤70%) favor island formation. Notably, Au and Pt also exhibit some island formation with low aspect ratios (Figure 4b and Figure 4c) for the strongest interaction strengths. In contrast, Cu (Figure 5c) and Ni (Figure 5d) show minimal island formation across the entire range of interaction strengths investigated.

While it is challenging to directly correlate the specific activation energy values used in this study with the diverse metal/substrate combinations found in the literature—requiring expensive ab initio calculations for each case—experimental observations show that transition and noble metals exhibit poor wetting on oxides surfaces[13] or 2D materials such as graphene[4] and $MoS_2$,[5] resulting in island formation and significant substrate exposure. An exception is Pd on $MoS_2$, which shows good wettability characteristic of strong substrate-metal interaction,[5] consistent with the behavior observed for strongly interacting substrates in our simulations.

Our simulated Au and Ag films on weakly interacting substrates align well with experimental observations. Atomic Force Microscopy (AFM) profiles of Au on $MoS_2$ show relatively flat surfaces with hole-like features among clusters, while Ag deposition results in island formation with smaller holes,[5] which are consistent with our simulation results for weak interaction conditions. This agreement between simulated and experimentally observed morphologies supports the validity of our approach. Given that both metal oxides and 2D materials are either currently present or expected to play increasingly important roles in future electronic device technologies, predicting these growth behaviours become very relevant.

For the strongly interacting substrate regime, Cu interconnect applications represent the most extensively studied case, where liner materials are commonly introduced between the dielectric oxide and Cu to improve wettability, reduce island formation and enhance surface coverage.[19] While our simulations may correspond to relatively large interaction strengths, due to the use of relatively low step descent barriers across the entire range, Figure S5 exhibits a good correspondence with experimental observations,[19] as the increased interaction strength results in reduced number of islands with lower aspect ratio and full coverage.

This agreement between our simulation results and experimental observations further support the validity of our approach and confirms the trends observed in Figure 5.

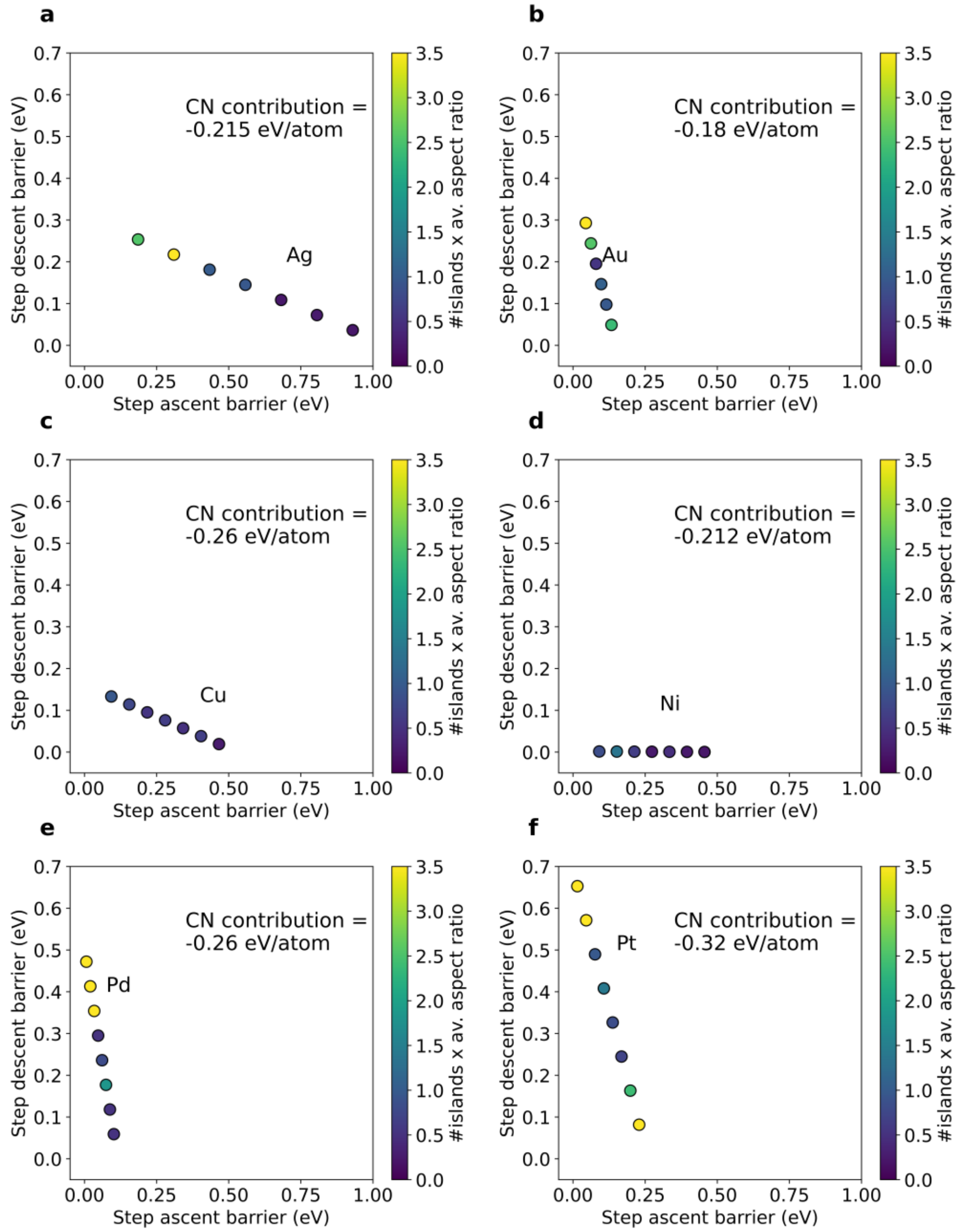

**Figure 5.** Heatmap representation of the island formation metric (island number × average aspect ratio) as a function of substrate interaction strength, with step ascent barrier (x-axis) and step descent barrier (y-axis) for (a) Ag, (b) Au, (c) Cu, (d) Ni, (e) Pd and (f) Pt. Consistent axis scales and color bars are maintained across all metals to facilitate comparison. High metric

values (yellow/green regions) indicate island formation, while low values (dark blue regions) correspond to flattened surface morphologies.

Figure 6 presents the atom fraction with a CN of 9 (Figure 6a) and 12 (Figure 6b), used as a proxy for defect density. The number of atoms with a given CN is shown for Ag (Figure S9a), Au (Figure S9b), Cu (Figure S9c), Ni (Figure S9d), Pd (Figure S9e) and Pt (Figure S9f). Ag, Cu and Ni exhibit a similar concentration of defect for substrate-metal interaction strengths ≥ 50%, likely due to their moderate CN contributions (-0.21 eV/atom for Ag and Ni, -0.26 eV/atom for Cu), low step descent and high step ascent barriers. Consequently, these metals are less sensitive to variations in substrate interaction strength (modifications of step ascent and step descent barriers). In contrast, Au, Pd and Pt—metals with the lowest step ascent barriers—exhibit pronounced morphological responses to changes in step ascent/descent barriers, consistent with trends observed in Figure 3 and 4. Ni exhibits the highest atom fraction with CN = 9 (0.62-0.70) and CN = 12 (0.90-0.95), showing a gradual increase with substrate interaction strength, followed by Cu and Ag. Notably, Ag shows comparable atom fraction with CN = 12 (Figure 6b) to those of Pd, Au and Pt at substrate interaction strengths ≥ 70%. For CN = 9, all metals converge to approximately 0.6 at substrate interaction strengths ≥ 90%, approaching homoepitaxial conditions. These results align with observations from the homoepitaxial case (Figure 1g). Au consistently shows the lowest atom fraction across all interaction strengths for CN = 12 (Figure 6b), although it converges with other metals for CN = 9 at substrate interaction strengths ≥ 90%. Pd and Pt show a similar behavior, but less pronounced. These results suggest that weak substrate interactions may enhance the catalytic activity of Au, Pd and Pt by increasing dangling bond density. Conversely, the fabrication of interconnects from these metals or their intermetallic compounds would require stronger substrate interactions to suppress defect formation.

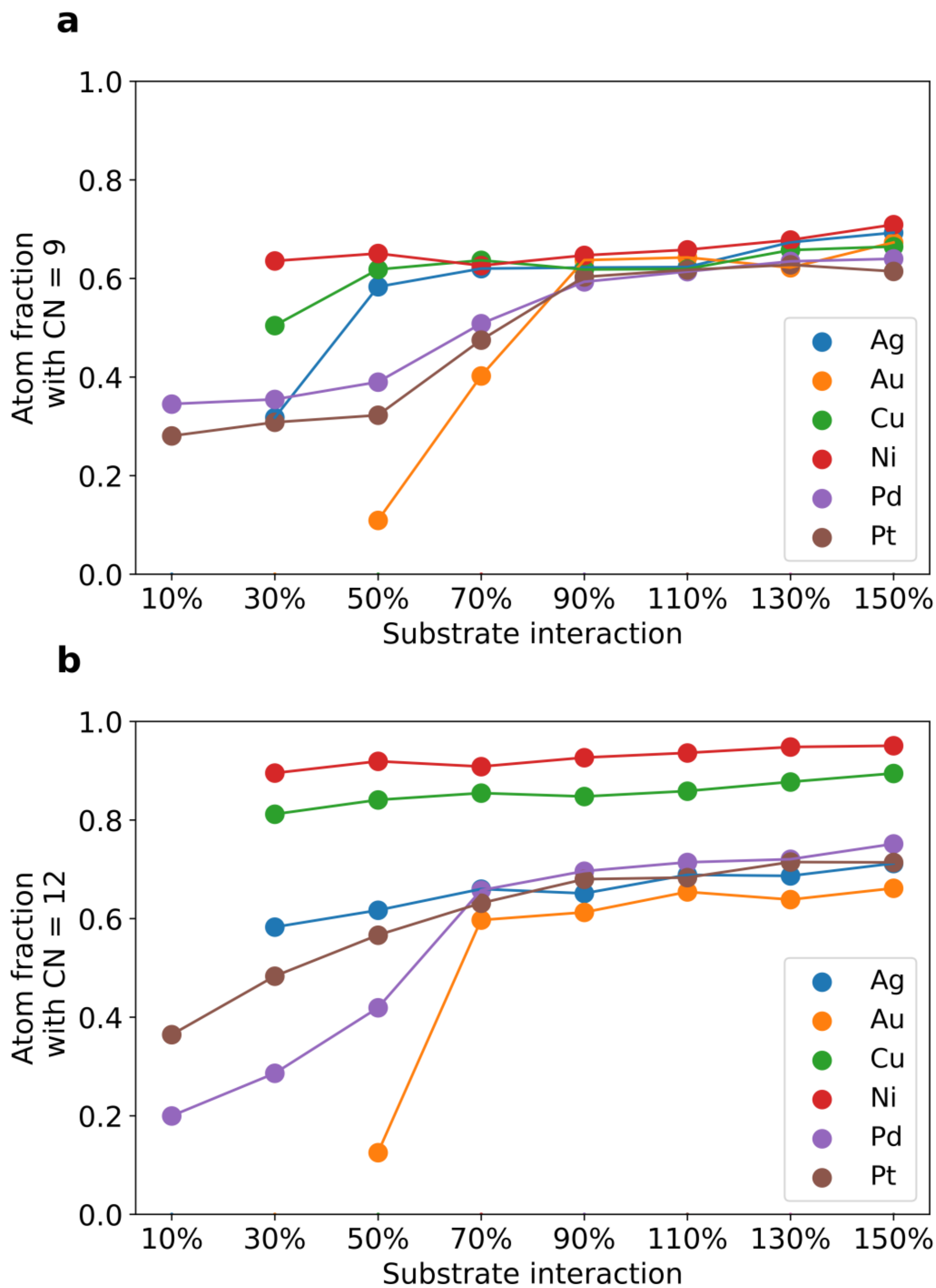

**Figure 6.** Coordination number (CN) analysis of atoms across varying substrate-metal interaction strengths as a proxy of defect density for Ag, Au, Cu, Ni, Pd, and Pt. (a) and (b) are the atom fraction with CN = 9 and CN = 12.

Figure 7 presents the layer-by-layer occupation rate, providing insight into film morphology when interpreted alongside the previous figures. Consistent with the trends observed in Figure 6, Ag, Cu and Ni show minimal changes in the occupation rate across layers across the entire range of substrate interaction strength. In contrast, Au, Pd and Pt show significant variations, with atomic distribution extending to higher layers under weaker interactions. For strong substrate interactions, the last layer with atoms present typically lies between the 6$^{th}$ and 8$^{th}$ layers. However, as the interaction weakens, atoms migrate to higher layers —up to the 11$^{th}$–14$^{th}$ layers —while the population in the lower layers diminishes, as shown for Au (Figure 7b), Pd (Figure 7e) and Pt (Figure 7f). Below 50% of the homepitaxial interaction strength, especially in the case of Au, metal detachment from the substrate may occur, suggesting a practical range of interaction ≥50%. Increasing substrate interaction strength beyond the homoepitaxial case provides limited benefit for enhancing lower-layer occupation rates or suppressing vertical growth, although we will discuss its relevance during thermal vacuum annealing in the following section. Notably, the first-layer occupation rate in Au, Pd and Pt increases markedly with stronger substrate interaction (see substrate exposure in Figure 3b), highlighting the effective role of interaction strength in anchoring the film to the substrate.

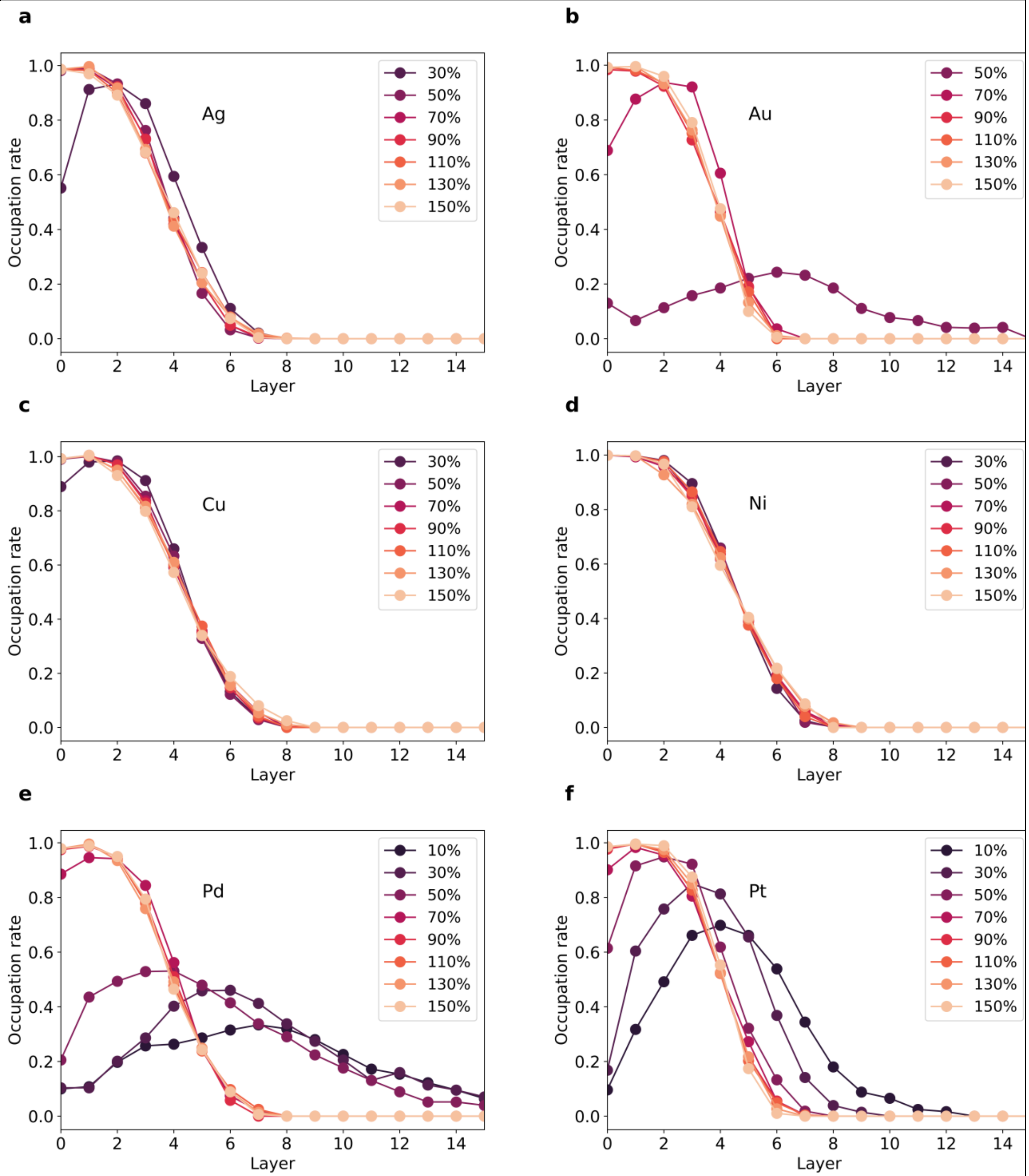

**Figure 7.** Layer-by-layer occupation rate as a function of substrate-metal interaction strengths for (a) Ag, (b) Au, (c) Cu, (d) Ni, (e) Pd, and (f) Pt. Color intensity indicates interaction strength, with lighter shades corresponding to stronger substrate-metal interactions.

**Impact of Thermal Vacuum Annealing on Metal Morphology**

Thermal vacuum annealing is a standard post-deposition process in thin film fabrication for interconnect, employed to modify morphology characteristics and reduce the defect density, with the aim of enhancing electrical and mechanical properties. However, it is important to note that while such improvements benefit interconnect applications, they may adversely affect catalytic functionality. We analyzed the effect of thermal annealing on the six metals (Ag, Au, Cu, Ni, Pd and Pt). Simulations were performed on films deposited onto substrates with three substrate-metal interaction strengths: weak (50%), homoepitaxial (100%) and strong (150%). Starting from the as-deposited simulated film, each system was annealed for 5 million kMC steps at 723 K, which is within an established thermal budget of 2h at that temperature for BEOL integration in 28-nm CMOS technology.[70] Due to metal-substrate specific energy barriers, the effective annealing durations for the 5 million kMC steps varied across systems (Table S1).

Figure 8 examines the impact of annealing on interconnect-relevant metrics previously presented in Figure 3, namely: RMS roughness, substrate exposure fraction relative to total area, and normalized maximum flat surface area relative to total area. For each metal, two sets of three bars are plotted to facilitate comparison between substrate interaction strengths and between as-deposited and post-annealed states. From left to right, bars correspond to 50%, 100% and 150% substrate interaction strengths, with solid color bars representing values for as-deposited metals and overlaid dashed bars representing the post-anneal values. Comparison among surface layers for as-deposited and post-annealed is provided in the Supporting Information for Ag (Figure S10), Au (Figure S11), Cu (Figure S12), Ni (Figure S13), Pd (Figure S14) and Pt (Figure S15).

Figure 8a demonstrates that Ag, Cu and Ni exhibit slight improvements in RMS roughness after annealing on homoepitaxial and strongly interacting substrate, although weakly interacting substrates result in minor roughness increases for Ag and Ni. Pd and Pt show clear deterioration in RMS roughness for weak and homoepitaxial cases, with only strongly interacting substrates yielding modest roughness reduction. Au presents a different behavior: weakly interacting and homoepitaxial substrates are insufficient to maintain metal adhesion, resulting in detachment, while annealing on strongly interacting substrates, though maintaining adhesion, increases film roughness. For the strongly interacting substrate case, Au forms an elongated island structure (Figure S11), resembling experimentally observed morphologies where Au forms extended structures on V substrates after annealing.[71]

Substrate exposure—or conversely, substrate coverage— represents another key metric for interconnect fabrication that must be evaluated to ensure the annealing process does not increase substrate exposure (Figure 8b). Ag, Cu and Ni exhibit minimal substrate exposure, with slight improvements for Ag on substrates ≥100% and a minor increase on weak substrates. In contrast, Au, Pd and Pt require strongly interacting substrates to maintain acceptable substrate exposure levels, as weaker substrates significantly increase exposure. Notably, as previously discussed, insufficient substrate interaction strength causes Au detachment, rendering it unsuitable for applications requiring annealing or high-temperature processing. Conversely, most materials exhibit increased normalized maximum flat surface area across the studied substrate interaction strengths, as shown in Figure 8c. Exceptions include Pd, which shows reduced flat surface area for interaction strengths ≤100%, and Au, which cannot maintain substrate adhesion. Pt on strongly interacting substrates shows the most significant improvement in normalized maximum surface area among all materials. The resulting planar morphology includes distributed holes, resembling the experimentally observed structure of Pt growth on $ZrO_2$, which consists of plateaus and holes.[72]

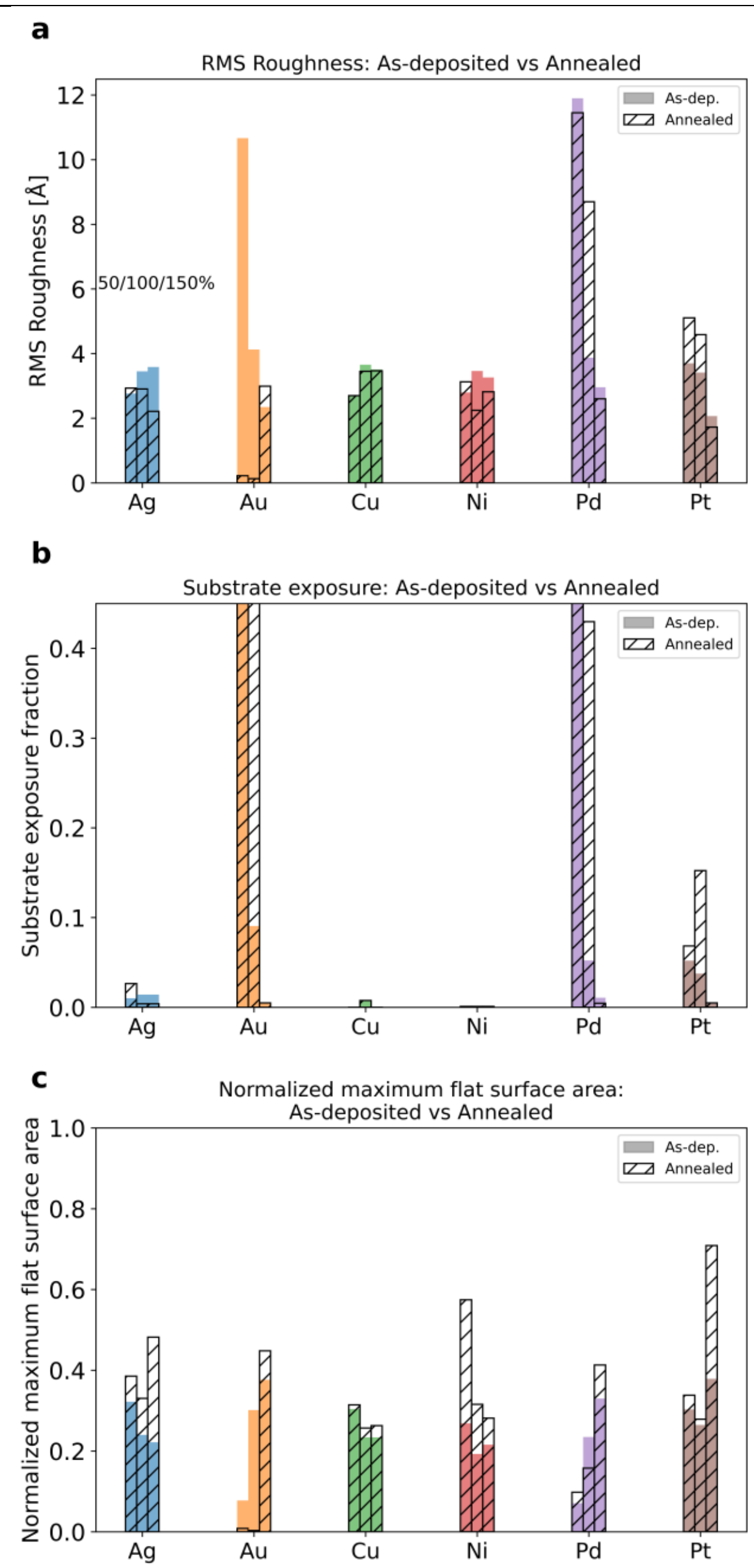

**Figure 8.** Effect of annealing at 723 K on Ag, Au, Cu, Ni, Pd and Pt films on metrics relevant for interconnect for three substrate-metal interaction strengths: weak (50%), homoepitaxial (100%) and strong (150%). For each metal, bars are ordered from left to right according to increasing interaction strength. Solid bars represent the as-deposited value, while hatched bars

indicate post-annealing values. (a) RMS roughness, (b) substrate exposure fraction relative to the total area and (d) normalized maximum flat surface area relative to the total area.

Figure 9 examines the impact of annealing on catalytically relevant metrics previously discussed in conjunction with Figure 4, specifically: island coverage fraction (Figure 9a), number of islands (Figure 9b) and mean aspect ratio of islands (Figure 9c). The bar format remains consistent with Figure 8, and surface layers for as-deposited and post-annealed conditions are presented in Figure S10-S15 for all six metals. These metrics should be interpreted together to evaluate film morphologies and their relationship to interconnect applications or catalytic activity, as significant differences exist between surfaces covered by dispersed small islands versus a single flat island. Figure 9a reveals that island coverage fraction generally increases following annealing, though certain cases exhibit substantial reduction. This behavior can be attributed to the decreased number of islands shown in Figure 9b and the reduced aspect ratios in Figure 9c. Only Pt on weakly interacting substrates results in an increase in island number, which is accompanied by aspect ratio reduction, indicating flatter island morphologies. Pt also exhibits flat surface formation with no island development on strongly interacting substrates, demonstrating a morphology approaching layer-by-layer growth mode (see Figure S15). The increased aspect ratio observed for Ag in the homoepitaxial case results from a single-particle that elevates the average value, though surface flattening is evident in Figure S10. Conversely, the aspect ratio increase for Pd stems from valley formation illustrated in Figure S14d. The annealing process increases substrate exposure by enlarging valleys present in the as-deposited film, therefore reducing the size of the base of the island and consequently increasing the aspect ratio. The combination of strongly interacting substrates with thermal vacuum annealing can be employed to promote flattening and island number reduction, promoting overall morphological smoothening.

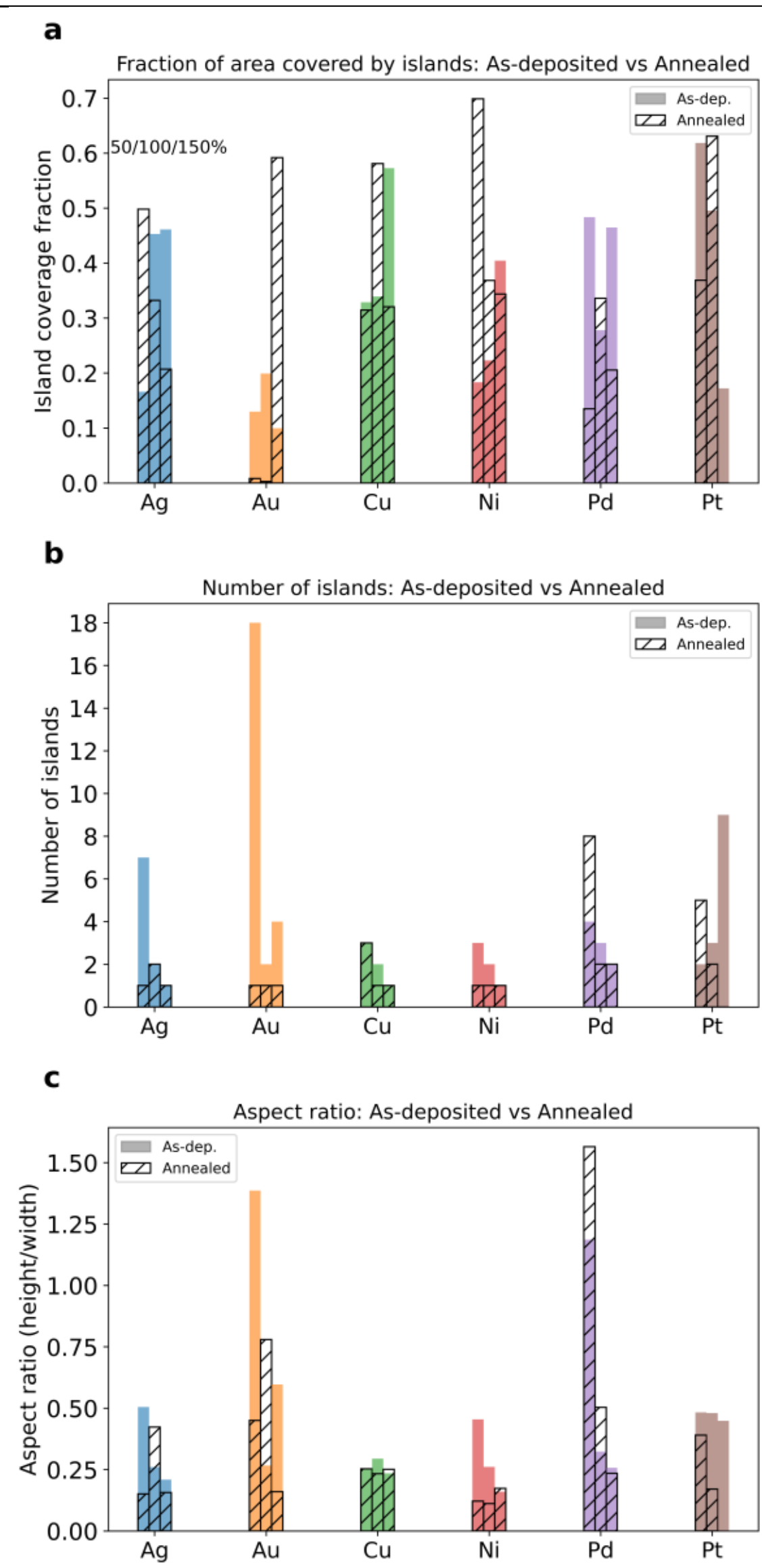

**Figure 9.** Effect of annealing at 723 K on Ag, Au, Cu, Ni, Pd and Pt films on metrics relevant for catalysis for three substrate-metal interaction strengths: weak (50%), homoepitaxial (100%) and strong (150%). For each metal, bars are ordered from left to right according to increasing interaction strength. Solid bars represent the as-deposited value, while hatched bars indicate post-annealing values. (a) island coverage fraction, (b) number of islands and (c) mean aspect ratio of islands (height-to-width).

Figure 10 presents the impact of thermal vacuum annealing on atom fractions with CN = 9 and CN = 12, which serve as proxies for defect density, as previously discussed for Figure 6. The bar format remains consistent with Figure 8 and 9, and surface layers for as-deposited and post-annealed conditions are presented in Figure S10-S15 for all six metals. Ag, Cu and Ni exhibit high atom fractions for both, CN = 9 and CN = 12 in as-deposited films (as previously discussed in Figure 6) and demonstrate clear improvements following annealing across all substrate interaction strengths. Their low step descent barriers combined with moderate CN contributions (-0.21 eV/atom for Ag and Ni, -0.26 eV/atom for Cu) facilitate thermal rearrangement with a preference for downward migration. Additionally, the CN contributions are sufficiently high to prevent atoms from detaching once they reach stable sites. Pd and Pt, although possessing some of the highest CN contributions (-0.26 eV/atom for Pd and -0.32 eV/atom for Pt) that favor stabilization of highly coordinated atoms, exhibit high step descent barriers that impede downward atomic migration. Consequently, these metals require strongly interacting substrates to promote downward migration and facilitate rearrangement toward more coordinated configurations. Au, with the lowest CN contribution (-0.18 eV/atom) and among the lowest step ascent barriers, detaches from substrate when interaction strength is insufficient. The low CN contribution facilitates the thermal arrangement, but results in less stable highly coordinated atoms compared to other metals. As a result, Au on strongly interacting substrates shows an increased atom fraction for CN = 9, but reduced atom fraction for CN = 12, likely due to plateau-type morphology formation following annealing. These results demonstrate that the combination of strongly interacting substrates with annealing processes provides an effective strategy for reducing defect density in metal films. This finding aligns with the well-established role of annealing in defect reduction during semiconductor fabrication, confirming that our approach captures this behavior while revealing the modulating effect of substrate interaction strength.[73]

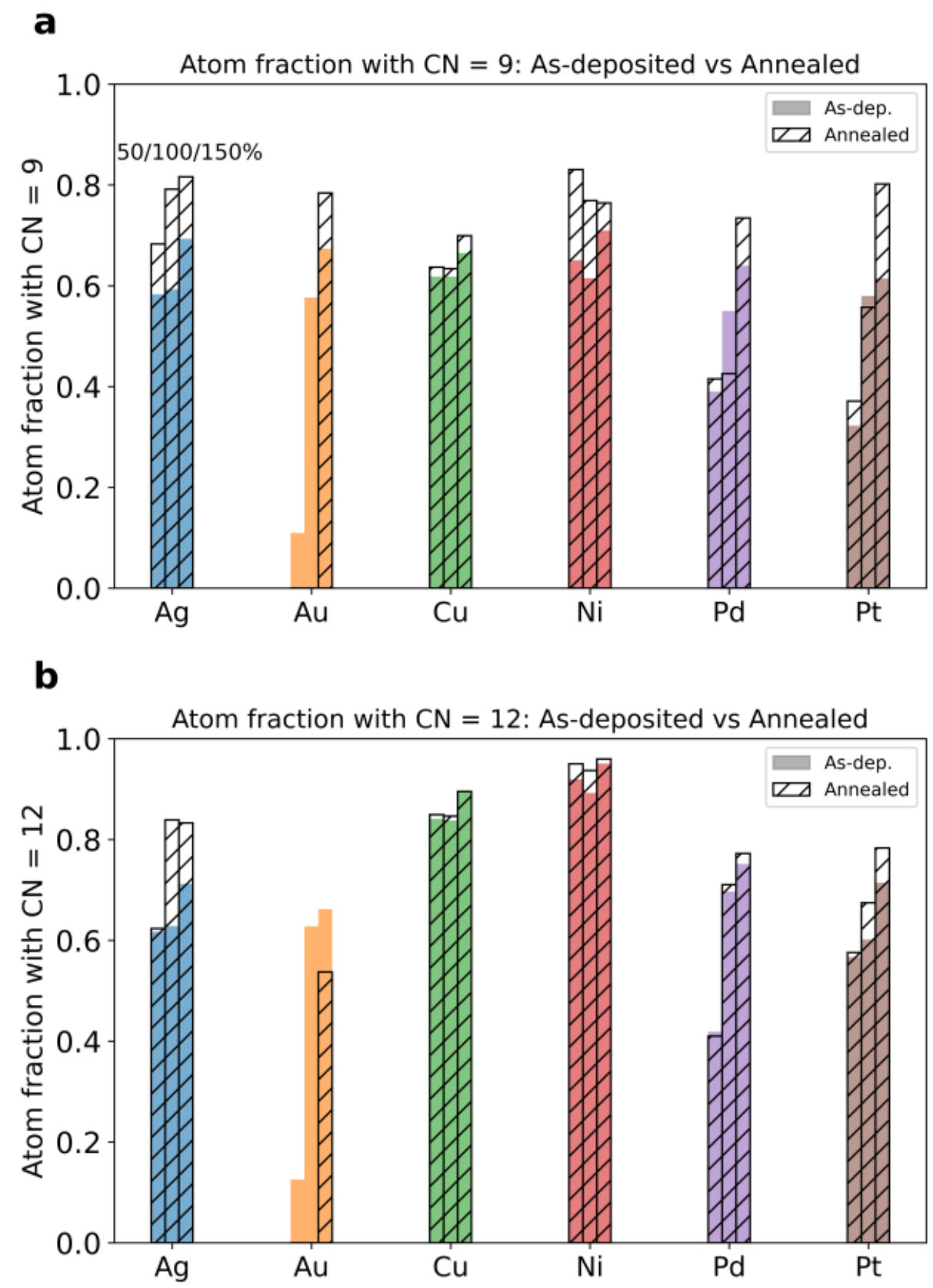

**Figure 10.** Coordination number (CN) analysis of atoms across varying substrate-metal interaction strengths (50%, 100% and 150%) for as-deposited (solid color bars) and post-annealed (dashed bars) films at 723 K as a proxy for defect density for Ag, Au, Cu, Ni, Pd, and Pt. Bars are ordered from left to right according to increasing interaction strength. (a) Atom fraction with CN = 9, (b) Atom fraction with CN = 12 (indicating fully coordinated atoms).

DISCUSSION AND OUTLOOK

Controlling film morphology is crucial for interconnect and catalysis applications. KMC simulations provide a powerful tool for screening and evaluating the growth and morphology of candidate metals for target applications such as catalysis and CMOS interconnect. This simulation framework supports the exploration of morphology control strategies through tuning the substrate-metal interaction strength and thermal vacuum annealing processes, aiming to achieve low-defect layer-by-layer growth films for interconnect applications or dispersions of small 3D clusters with high densities of undercoordinated atoms for catalytic purposes.

The agreement between our simulation results and various experimental observations across different conditions (homoepitaxial growth, varying substrate interactions and thermal annealing) supports the validity of our model, which captures fundamental growth tendencies under realistic experimental conditions.

Simulations of the homoepitaxial case reveal the critical role of interplay of the upwards and downwards activation energy for metal migration in determining film morphologies, enabling classification of the metals into two distinct groups. Au, Pd and Pt, exhibit the lowest step ascent energy barriers and highest step descent energy barriers, with significant differences in the CN contribution to activation energies—Au showing the lowest CN contribution and Pt the highest. These three metals demonstrate the largest substrate exposure among the materials studied, with Au and Pd exhibiting the highest roughness values, with Pt showing the lowest roughness. In contrast, Ag, Cu and Ni show significantly less substrate exposure and lower roughness values, that are comparable to Pt. The complex interplay between step ascent and step descent barriers and CN contributions to activation energies is challenging to evaluate without these long timescale kinetic simulations and has significant relevance when using substrate-metal interaction strength to tune the target metal morphology and employ thermal vacuum annealing conditions.

Variations in substrate-metal interaction strength, modeled by facilitating upward migration while hindering downward migration for weakly interacting substrate and the opposite for strongly ones, exert distinctly different impacts on the two metal groups. Au, Pd and Pt demonstrate the highest sensitivity to substrate interaction strength variations (in that order), showing how the transition from weak to strong interacting substrates decreases the RMS roughness, substrate exposure, island number and island aspect ratios, while simultaneously increasing flat surface areas and the atomic CN (indicating reduced defect density). This demonstrates that substrate interaction strength can be effectively employed to mitigate island formation and promote layer-by-layer growth modes, although it cannot achieve large flat surface areas, which remain < 40% across all substrate interaction strengths. Notably, the values exhibited by Au, Pd and Pt for different metrics converge toward those of Ag, Cu and Ni with increasing substrate interaction strength: RMS roughness decreases from ~9-11 Å to ~2.5–3.5 Å (for films with an average thickness of 10 Å), substrate exposure fraction decreases from 0.3–0.76 on weakly interacting substrates (≤50%) to 0.005–0.02 at 90% interaction strength, normalized maximum flat surface areas increase from ≤0.2 on weakly interacting substrates to 0.22–0.4, and island number and aspect ratios decrease to 1-3 islands and 0.3, respectively. In summary, as-deposited Au, Pd, Pt and lastly Ag exhibit the most favorable properties for catalytic applications and are of course studied as supported nanoparticles for catalysis. While these metals are expensive, they can be alloyed with cheaper metals to achieve dispersions of small three-dimensional clusters, making them viable for practical catalytic applications, which is also a topic of high interest. No as-deposited metals display structure morphology relevant for interconnect, so we therefore need to introduce the vacuum anneal.

Our simulations demonstrate that thermal vacuum annealing, a standard post-deposition process in thin film fabrication, can clearly improve interconnect-relevant metrics when the substrate interactions are sufficiently strong relative to the homoepitaxial case. Conversely, weak substrate interactions during thermal annealing can deteriorate RMS roughness, increase substrate exposure or even increase island aspect ratio. In extreme cases, such as Au, weak interaction may be insufficient to maintain adhesion to the substrate. In contrast, sufficiently strong substrate interactions enable thermal vacuum annealing to reduce RMS roughness, significantly increase normalized maximum flat surface areas, reduce island numbers through merging, and decrease island aspect ratios. Additionally, annealing increases atomic coordination numbers, thereby reducing defect density. Consequently, the annealing process promotes film smoothening while simultaneously reducing defect density, favoring metrics relevant for interconnect. We note that the substrate-metal interaction energy can be evaluated from moderately sized DFT calculations as described in references.[18,68]

For interconnect applications, where alloying is an increasingly important strategy for addressing the challenges of miniaturization, our results support the search for Cu replacements and the development of alloy components. The ongoing miniaturization of electronic devices presents significant challenges for Cu interconnect deposition due to Cu's tendency to form 3D clusters at small scales. Our findings highlight how specific metals offer morphological benefits when used as alloying components and identifying potential candidates based on both structural and electrical properties.

While Ag and Ni demonstrate low RMS roughness and substrate exposure fractions after annealing, and Pd and Pt can achieve similar or superior values on strongly interacting substrates, the selection of alloying elements must also consider electrical performance. The residual resistivity of binary alloys generally increases with impurity concentration according to Nordheim's rule, which is a limiting factor.[74] For instance, starting from Cu as the primary material for interconnect fabrication, Ag, Pd and Au are potentially good candidates for alloying due to their relatively small contribution to residual resistivity, while alloys incorporating Ni or Pt typically exhibit large residual resistivity, potentially eliminating these as alloying elements based on this criterion alone.[75] Furthermore, Pd and Pt exhibit a weaker film thickness dependence of resistivity compared to Cu,[76] making them interesting candidates for alloying with cheaper materials to relax cost constraints, or for critical applications where cost is less restrictive.

Additionally, Ag, Pd and Au show flatter surfaces with minimal substrate exposure and reduced RMS roughness after annealing on strongly interacting substrates (Figure S10f, S11f and S14f) compared to Cu (Figure S12f), suggesting potential benefits in terms of reduced defects and migration resistance. Ni, while presenting higher residual resistivity in alloys, offers excellent morphological features (low RMS roughness, reduced defect density, minimal substrate exposure, relatively large flat surfaces post-annealing) and beneficial electronic properties for scaling, such as shorter electron mean free path than Cu and Ag, which reduces scattering at smaller dimensions,[77] a factor already leveraged in NiCo[78] and NiAl[79] alloys. Consequently, these materials represent promising candidates for alloy components, with the choice depending on the balance required between morphological control, electrical resistivity and cost. Although Ag, Pd and Pt are significantly more expensive than Cu, representing a major limitation for their use as single elemental metals, they may be effectively combined with cheaper metals in alloy systems to leverage their beneficial morphological and, for Pd/Pt, potentially favorable resistivity scaling properties while maintaining cost-effectiveness.

A major challenge in employing the present simulation approach is the computational cost required to obtain realistic outputs. The time required for DFT calculations to obtain a limited set of energy barriers for the rate constants is a serious limitation, while obtaining the necessary barriers in binary alloys can be computationally impractical, and even worse for ternary systems, due to the enormous number of possible atomic configurations and possible migrations involved.

A potential solution to address this challenge involves using DFT-trained machine learning interatomic potentials, which enable significantly faster determination of required barriers and metal-substrate interaction strengths (in the range of minutes) compared to standard DFT calculations.[68] This approach would allow on-the-fly calculation of activation barriers for kMC simulations, paving the way to simulate much more complex systems with enhanced accuracy.

## CONCLUSIONS

We employed kinetic Monte Carlo simulations to investigate the influence of substrate-metal interaction strength and thermal vacuum annealing on the morphology of six fcc metals (Ag, Au, Cu, Ni, Pd, Pt). Our simulations reveal distinct morphological responses to varying interaction strength (modeled by modifying the step ascent and descent barriers relative to the homoepitaxial case), allowing classification into two groups: (i) Au, Pd, Pt (high sensitivity to interaction strength, favoring island formation on weak substrates) and (ii) Ag, Cu, Ni (low sensitivity to interaction strength, showing smoother films). Substrate interaction strength can be employed to control the transition from 3D island to 2D layer-by-layer growth, although achieving large flat surface areas remains challenging. Thermal vacuum annealing significantly improves interconnect-relevant metrics (such as roughness, defect density, normalized maximum flat surface areas, island numbers and island aspect ratios) when employed with sufficiently strong substrate interactions. These results provide insights for alloy design strategies tailored to interconnect and catalysis applications.

## ASSOCIATED CONTENT

**Supporting Information**.

The following files are available free of charge.

Supporting Information showing: key metrics, surface layer morphologies resulting from deposition and after annealing, coordination number data and annealing time for each metal. (PDF)

Supporting information: GIF of a typical fcc metal deposition (Pt) and GIF of a typical thermal vacuum annealing of Pt. Link.

AUTHOR INFORMATION

**Corresponding Author**

Samuel Aldana - Tyndall National Institute, University College Cork, Lee Maltings, Dyke Parade, Cork T12 R5CP, Ireland. https://orcid.org/0000-0002-1719-605X

E-mail: samuel.delgado@tyndall.ie

Michael Nolan - Tyndall National Institute, University College Cork, Lee Maltings, Dyke Parade, Cork T12 R5CP, Ireland. https://orcid.org/0000-0002-5224-8580

E-mail: michael.nolan@tyndall.ie

**Author Contributions**

S. A. developed and conducted the kMC simulations. S.A. and M. N. analyzed the data. M. N. supervised the project. The manuscript was written through contributions of all authors. All authors have given approval to the final version of the manuscript.

**Funding Sources**

The ASCENT+ Access to European Infrastructure for Nanoelectronics Program, funded through the EU Horizon Europe programme, grant no 871130 supported MN and SD. MN was supported through the Science Foundation Ireland SFI-NSF China Partnership program, grant number 17/NSFC/5279.

**Conflict of interest**

There are no conflicts of interest to declare.

**Data availability**

The code for the kMC simulator can be found at Github.

References

(1) Jamnig, A.; Pliatsikas, N.; Konpan, M.; Lu, J.; Kehagias, T.; Kotanidis, A. N.; Kalfagiannis, N.; Bellas, D. V.; Lidorikis, E.; Kovac, J.; Abadias, G.; Petrov, I.; Greene, J. E.; Sarakinos, K. 3D-to-2D Morphology Manipulation of Sputter-Deposited Nanoscale Silver Films on Weakly Interacting Substrates via Selective Nitrogen Deployment for Multifunctional Metal Contacts. *ACS Appl. Nano Mater.* **2020**, *3* (5), 4728–4738. https://doi.org/10.1021/acsanm.0c00736.

(2) Mueller, T.; Xia, F.; Avouris, P. Graphene Photodetectors for High-Speed Optical Communications. *Nat. Photonics* **2010**, *4* (5), 297–301. https://doi.org/10.1038/nphoton.2010.40.

(3) Xu, Y.; Hsieh, C.-Y.; Wu, L.; Ang, L. K. Two-Dimensional Transition Metal Dichalcogenides Mediated Long Range Surface Plasmon Resonance Biosensors. *J. Phys. Appl. Phys.* **2019**, *52* (6), 065101. https://doi.org/10.1088/1361-6463/aaf0f7.

(4) Liu, X.; Han, Y.; Evans, J. W.; Engstfeld, A. K.; Behm, R. J.; Tringides, M. C.; Hupalo, M.; Lin, H.-Q.; Huang, L.; Ho, K.-M.; Appy, D.; Thiel, P. A.; Wang, C.-Z. Growth Morphology and Properties of Metals on Graphene. *Prog. Surf. Sci.* **2015**, *90* (4), 397–443. https://doi.org/10.1016/j.progsurf.2015.07.001.

(5) Gong, C.; Huang, C.; Miller, J.; Cheng, L.; Hao, Y.; Cobden, D.; Kim, J.; Ruoff, R. S.; Wallace, R. M.; Cho, K.; Xu, X.; Chabal, Y. J. Metal Contacts on Physical Vapor Deposited Monolayer $MoS_2$. *ACS Nano* **2013**, *7* (12), 11350–11357. https://doi.org/10.1021/nn4052138.

(6) Kim, J.-S.; Kim, J.; Yang, D.-J.; Shim, J.; Hu, L.; Lee, C.; Kim, J.; Kim, S. W. Addressing Interconnect Challenges for Enhanced Computing Performance. *Science* **2024**, *386* (6727), eadk6189. https://doi.org/10.1126/science.adk6189.

(7) Grillo, F.; Moulijn, J. A.; Kreutzer, M. T.; Van Ommen, J. R. Nanoparticle Sintering in Atomic Layer Deposition of Supported Catalysts: Kinetic Modeling of the Size Distribution. *Catal. Today* **2018**, *316*, 51–61. https://doi.org/10.1016/j.cattod.2018.02.020.
(8) Dimian, A. C.; Bildea, C. S.; Kiss, A. A. Process Intensification. In *Computer Aided Chemical Engineering*; Elsevier, 2014; Vol. 35, pp 397–448. https://doi.org/10.1016/B978-0-444-62700-1.00010-3.
(9) Kandel, D.; Kaxiras, E. The Surfactant Effect in Semiconductor Thin-Film Growth. In *Solid State Physics*; Elsevier, 2000; Vol. 54, pp 219–262. https://doi.org/10.1016/S0081-1947(08)60249-0.
(10) Roldan Cuenya, B.; Behafarid, F. Nanocatalysis: Size- and Shape-Dependent Chemisorption and Catalytic Reactivity. *Surf. Sci. Rep.* **2015**, *70* (2), 135–187. https://doi.org/10.1016/j.surfrep.2015.01.001.
(11) Liu, J. Catalysis by Supported Single Metal Atoms. *ACS Catal.* **2017**, *7* (1), 34–59. https://doi.org/10.1021/acscatal.6b01534.
(12) Van Bui, H.; Grillo, F.; Kulkarni, S. S.; Bevaart, R.; Van Thang, N.; Van Der Linden, B.; Moulijn, J. A.; Makkee, M.; Kreutzer, M. T.; Van Ommen, J. R. Low-Temperature Atomic Layer Deposition Delivers More Active and Stable Pt-Based Catalysts. *Nanoscale* **2017**, *9* (30), 10802–10810. https://doi.org/10.1039/C7NR02984E.
(13) Campbell, C. T. Metal Films and Particles on Oxide Surfaces: Structural, Electronic and Chemisorptive Properties. *J. Chem. Soc. Faraday Trans.* **1996**, *92* (9), 1435. https://doi.org/10.1039/ft9969201435.
(14) Schlögl, R. Heterogeneous Catalysis. *Angew. Chem. Int. Ed.* **2015**, *54* (11), 3465–3520. https://doi.org/10.1002/anie.201410738.
(15) Mackus, A. J. M.; Weber, M. J.; Thissen, N. F. W.; Garcia-Alonso, D.; Vervuurt, R. H. J.; Assali, S.; Bol, A. A.; Verheijen, M. A.; Kessels, W. M. M. Atomic Layer Deposition of Pd and Pt Nanoparticles for Catalysis: On the Mechanisms of Nanoparticle Formation. *Nanotechnology* **2016**, *27* (3), 034001. https://doi.org/10.1088/0957-4484/27/3/034001.
(16) Lü, B.; Almyras, G. A.; Gervilla, V.; Greene, J. E.; Sarakinos, K. Formation and Morphological Evolution of Self-Similar 3D Nanostructures on Weakly Interacting Substrates. *Phys. Rev. Mater.* **2018**, *2* (6), 063401. https://doi.org/10.1103/PhysRevMaterials.2.063401.
(17) Zhao, G.; Wang, W.; Bae, T.-S.; Lee, S.-G.; Mun, C.; Lee, S.; Yu, H.; Lee, G.-H.; Song, M.; Yun, J. Stable Ultrathin Partially Oxidized Copper Film Electrode for Highly Efficient Flexible Solar Cells. *Nat. Commun.* **2015**, *6* (1), 8830. https://doi.org/10.1038/ncomms9830.
(18) Nies, C.-L.; Natarajan, S. K.; Nolan, M. Control of the Cu Morphology on Ru-Passivated and Ru-Doped TaN Surfaces – Promoting Growth of 2D Conducting Copper for CMOS Interconnects. *Chem. Sci.* **2022**, *13* (3), 713–725. https://doi.org/10.1039/D1SC04708F.
(19) Kim, H.; Naito, Y.; Koseki, T.; Ohba, T.; Ohta, T.; Kojima, Y.; Sato, H.; Shimogaki, Y. Material Consideration on Ta, Mo, Ru, and Os as Glue Layer for Ultra Large Scale Integration Cu Interconnects. *Jpn. J. Appl. Phys.* **2006**, *45* (4R), 2497. https://doi.org/10.1143/JJAP.45.2497.
(20) Zhu, W.; Yogeesh, M. N.; Yang, S.; Aldave, S. H.; Kim, J.-S.; Sonde, S.; Tao, L.; Lu, N.; Akinwande, D. Flexible Black Phosphorus Ambipolar Transistors, Circuits and AM Demodulator. *Nano Lett.* **2015**, *15* (3), 1883–1890. https://doi.org/10.1021/nl5047329.

(21) Late, D. J.; Huang, Y.-K.; Liu, B.; Acharya, J.; Shirodkar, S. N.; Luo, J.; Yan, A.; Charles, D.; Waghmare, U. V.; Dravid, V. P.; Rao, C. N. R. Sensing Behavior of Atomically Thin-Layered $MoS_2$ Transistors. *ACS Nano* **2013**, *7* (6), 4879–4891. https://doi.org/10.1021/nn400026u.
(22) Lopez-Sanchez, O.; Lembke, D.; Kayci, M.; Radenovic, A.; Kis, A. Ultrasensitive Photodetectors Based on Monolayer MoS2. *Nat. Nanotechnol.* **2013**, *8* (7), 497–501. https://doi.org/10.1038/nnano.2013.100.
(23) Zhou, H.; Qiu, C.; Liu, Z.; Yang, H.; Hu, L.; Liu, J.; Yang, H.; Gu, C.; Sun, L. Thickness-Dependent Morphologies of Gold on *N* -Layer Graphenes. *J. Am. Chem. Soc.* **2010**, *132* (3), 944–946. https://doi.org/10.1021/ja909228n.
(24) N'Diaye, A. T.; Bleikamp, S.; Feibelman, P. J.; Michely, T. Two-Dimensional Ir Cluster Lattice on a Graphene Moiré on Ir(111). *Phys. Rev. Lett.* **2006**, *97* (21), 215501. https://doi.org/10.1103/PhysRevLett.97.215501.
(25) Han, Y.; Engstfeld, A. K.; Behm, R. J.; Evans, J. W. Atomistic Modeling of the Directed-Assembly of Bimetallic Pt-Ru Nanoclusters on Ru(0001)-Supported Monolayer Graphene. *J. Chem. Phys.* **2013**, *138* (13), 134703. https://doi.org/10.1063/1.4798348.
(26) Hutchings, G. J. Heterogeneous Gold Catalysis. *ACS Cent. Sci.* **2018**, *4* (9), 1095–1101. https://doi.org/10.1021/acscentsci.8b00306.
(27) Cui, Y.; Huang, K.; Nilius, N.; Freund, H.-J. Charge Competition with Oxygen Molecules Determines the Growth of Gold Particles on Doped CaO Films. *Faraday Discuss.* **2013**, *162*, 153. https://doi.org/10.1039/c3fd20130a.
(28) Haruta, M. Size- and Support-Dependency in the Catalysis of Gold. *Catal. Today* **1997**, *36* (1), 153–166. https://doi.org/10.1016/S0920-5861(96)00208-8.
(29) Giangregorio, M. M.; Losurdo, M.; Bianco, G. V.; Dilonardo, E.; Capezzuto, P.; Bruno, G. Synthesis and Characterization of Plasmon Resonant Gold Nanoparticles and Graphene for Photovoltaics. *Mater. Sci. Eng. B* **2013**, *178* (9), 559–567. https://doi.org/10.1016/j.mseb.2012.10.034.
(30) Zeng, S.; Yong, K.-T.; Roy, I.; Dinh, X.-Q.; Yu, X.; Luan, F. A Review on Functionalized Gold Nanoparticles for Biosensing Applications. *Plasmonics* **2011**, *6* (3), 491–506. https://doi.org/10.1007/s11468-011-9228-1.
(31) Fang, H.; Wu, Y.; Zhao, J.; Zhu, J. Silver Catalysis in the Fabrication of Silicon Nanowire Arrays. *Nanotechnology* **2006**, *17* (15), 3768–3774. https://doi.org/10.1088/0957-4484/17/15/026.
(32) Larciprete, M. C.; Albertoni, A.; Belardini, A.; Leahu, G.; Li Voti, R.; Mura, F.; Sibilia, C.; Nefedov, I.; Anoshkin, I. V.; Kauppinen, E. I.; Nasibulin, A. G. Infrared Properties of Randomly Oriented Silver Nanowires. *J. Appl. Phys.* **2012**, *112* (8), 083503. https://doi.org/10.1063/1.4759374.
(33) Hensel, B.; Grasso, G.; Flükiger, R. Limits to the Critical Transport Current in Superconducting (Bi,Pb ) 2 Sr 2 Ca 2 Cu 3 O 10 Silver-Sheathed Tapes: The Railway-Switch Model. *Phys. Rev. B* **1995**, *51* (21), 15456–15473. https://doi.org/10.1103/PhysRevB.51.15456.
(34) Duan, Y.; Rani, S.; Zhang, Y.; Ni, C.; Newberg, J. T.; Teplyakov, A. V. Silver Deposition onto Modified Silicon Substrates. *J. Phys. Chem. C* **2017**, *121* (13), 7240–7247. https://doi.org/10.1021/acs.jpcc.6b12896.

(35) Aldana, S.; Nies, C.-L.; Nolan, M. Control of Cu Morphology on TaN Barrier and Combined Ru-TaN Barrier/Liner Substrates for Nanoscale Interconnects from Atomistic Kinetic Monte Carlo Simulations.
(36) Soulié, J.-P.; Sankaran, K.; Van Troeye, B.; Leśniewska, A.; Varela Pedreira, O.; Oprins, H.; Delie, G.; Fleischmann, C.; Boakes, L.; Rolin, C.; Ragnarsson, L.-Å.; Croes, K.; Park, S.; Swerts, J.; Pourtois, G.; Tőkei, Z.; Adelmann, C. Selecting Alternative Metals for Advanced Interconnects. *J. Appl. Phys.* **2024**, *136* (17), 171101. https://doi.org/10.1063/5.0224027.
(37) Gall, D.; Cha, J. J.; Chen, Z.; Han, H.-J.; Hinkle, C.; Robinson, J. A.; Sundararaman, R.; Torsi, R. Materials for Interconnects. *MRS Bull.* **2021**, *46* (10), 959–966. https://doi.org/10.1557/s43577-021-00192-3.
(38) Koike, J.; Kuge, T.; Chen, L.; Yahagi, M. Intermetallic Compounds For Interconnect Metal Beyond 3 Nm Node. In *2021 IEEE International Interconnect Technology Conference (IITC)*; IEEE: Kyoto, Japan, 2021; pp 1–3. https://doi.org/10.1109/IITC51362.2021.9537364.
(39) Han, B.; Wu, J.; Zhou, C.; Chen, B.; Gordon, R.; Lei, X.; Roberts, D. A.; Cheng, H. First-Principles Simulations of Conditions of Enhanced Adhesion Between Copper and TaN(111) Surfaces Using a Variety of Metallic Glue Materials. *Angew. Chem. Int. Ed.* **2010**, *49* (1), 148–152. https://doi.org/10.1002/anie.200905360.
(40) Kondati Natarajan, S.; Nies, C.-L.; Nolan, M. Ru Passivated and Ru Doped ε-TaN Surfaces as a Combined Barrier and Liner Material for Copper Interconnects: A First Principles Study. *J. Mater. Chem. C* **2019**, *7* (26), 7959–7973. https://doi.org/10.1039/C8TC06118A.
(41) Kondati Natarajan, S.; Nies, C.-L.; Nolan, M. The Role of Ru Passivation and Doping on the Barrier and Seed Layer Properties of Ru-Modified TaN for Copper Interconnects. *J. Chem. Phys.* **2020**, *152* (14), 144701. https://doi.org/10.1063/5.0003852.
(42) Mao, R.; Kong, B. D.; Gong, C.; Xu, S.; Jayasekera, T.; Cho, K.; Kim, K. W. First-Principles Calculation of Thermal Transport in Metal/Graphene Systems. *Phys. Rev. B* **2013**, *87* (16), 165410. https://doi.org/10.1103/PhysRevB.87.165410.
(43) Kim, S. Y.; Lee, I.-H.; Jun, S. Transition-Pathway Models of Atomic Diffusion on Fcc Metal Surfaces. I. Flat Surfaces. *Phys. Rev. B* **2007**, *76* (24), 245407. https://doi.org/10.1103/PhysRevB.76.245407.
(44) Kim, S. Y.; Lee, I.-H.; Jun, S. Transition-Pathway Models of Atomic Diffusion on Fcc Metal Surfaces. II. Stepped Surfaces. *Phys. Rev. B* **2007**, *76* (24), 245408. https://doi.org/10.1103/PhysRevB.76.245408.
(45) Hayat, S. S.; Alcántara Ortigoza, M.; Choudhry, M. A.; Rahman, T. S. Diffusion of the Cu Monomer and Dimer on Ag(111): Molecular Dynamics Simulations and Density Functional Theory Calculations. *Phys. Rev. B* **2010**, *82* (8), 085405. https://doi.org/10.1103/PhysRevB.82.085405.
(46) Sun, G. S.; Jónsson, H. Kinetic Monte Carlo Simulation Studies of the Shape of Islands on Close-Packed Surfaces. *J. Electrochem. Soc.* **2022**, *169* (10), 102503. https://doi.org/10.1149/1945-7111/ac96ac.
(47) Aldana, S.; Wang, L.; Spiridon, I. A.; Zhang, H. Understanding Substrate Effects on 2D $MoS_2$ Growth: A Kinetic Monte Carlo Approach. *Adv. Mater. Interfaces* **2024**, 2400209. https://doi.org/10.1002/admi.202400209.

(48) Aldana, S.; Jadwiszczak, J.; Zhang, H. On the Switching Mechanism and Optimisation of Ion Irradiation Enabled 2D MoS2 Memristors. *Nanoscale* **2023**, *15* (13). https://doi.org/10.1039/d2nr06810a.
(49) Aldana, S.; Garciá-Fernández, P.; Romero-Zaliz, R.; González, M. B.; Jiménez-Molinos, F.; Gómez-Campos, F.; Campabadal, F.; Roldán, J. B. Resistive Switching in HfO2 Based Valence Change Memories, a Comprehensive 3D Kinetic Monte Carlo Approach. *J. Phys. Appl. Phys.* **2020**, *53* (22). https://doi.org/10.1088/1361-6463/ab7bb6.
(50) Battaile, C. C. The Kinetic Monte Carlo Method: Foundation, Implementation, and Application. *Comput. Methods Appl. Mech. Eng.* **2008**, *197* (41–42), 3386–3398. https://doi.org/10.1016/j.cma.2008.03.010.
(51) Kim, S.; An, H.; Oh, S.; Jung, J.; Kim, B.; Nam, S. K.; Han, S. Atomistic Kinetic Monte Carlo Simulation on Atomic Layer Deposition of TiN Thin Film. *Comput. Mater. Sci.* **2022**, *213*. https://doi.org/10.1016/j.commatsci.2022.111620.
(52) Prud'homme, N.; Constantoudis, V.; Turgambaeva, A. E.; Krisyuk, V. V.; Samélor, D.; Senocq, F.; Vahlas, C. Chemical Vapor Deposition of Cu Films from Copper(I) Cyclopentadienyl Triethylphophine: Precursor Characteristics and Interplay between Growth Parameters and Films Morphology. *Thin Solid Films* **2020**, *701*, 137967. https://doi.org/10.1016/j.tsf.2020.137967.
(53) Abrams, C. F.; Graves, D. B. Cu Sputtering and Deposition by Off-Normal, near-Threshold Cu+ Bombardment: Molecular Dynamics Simulations. *J. Appl. Phys.* **1999**, *86* (4), 2263–2267. https://doi.org/10.1063/1.371040.
(54) Voter, A. F.; Los Alamos Natl Lab, L. A. N. M. U. S. A. INTRODUCTION TO THE KINETIC MONTE CARLO METHOD; Springer: Erice, ITALY, 2004; Vol. 235, pp 1–23.
(55) Fichthorn, K. A.; Lin, Y. A Local Superbasin Kinetic Monte Carlo Method. *J. Chem. Phys.* **2013**, *138* (16), 164104. https://doi.org/10.1063/1.4801869.
(56) Chatterjee, A.; Vlachos, D. G. An Overview of Spatial Microscopic and Accelerated Kinetic Monte Carlo Methods. *J. Comput.-Aided Mater. Des.* **2007**, *14* (2), 253–308. https://doi.org/10.1007/s10820-006-9042-9.
(57) Ong, S. P.; Richards, W. D.; Jain, A.; Hautier, G.; Kocher, M.; Cholia, S.; Gunter, D.; Chevrier, V. L.; Persson, K. A.; Ceder, G. Python Materials Genomics (Pymatgen): A Robust, Open-Source Python Library for Materials Analysis. *Comput. Mater. Sci.* **2013**, *68*, 314–319. https://doi.org/10.1016/j.commatsci.2012.10.028.
(58) *Spglib: a software library for crystal symmetry search*. https://www.tandfonline.com/doi/epdf/10.1080/27660400.2024.2384822?needAccess=true (accessed 2025-04-08).
(59) Jain, A.; Ong, S. P.; Hautier, G.; Chen, W.; Richards, W. D.; Dacek, S.; Cholia, S.; Gunter, D.; Skinner, D.; Ceder, G.; Persson, K. A. Commentary: The Materials Project: A Materials Genome Approach to Accelerating Materials Innovation. *APL Mater.* **2013**, *1* (1), 011002. https://doi.org/10.1063/1.4812323.
(60) Ong, S. P.; Cholia, S.; Jain, A.; Brafman, M.; Gunter, D.; Ceder, G.; Persson, K. A. The Materials Application Programming Interface (API): A Simple, Flexible and Efficient API for Materials Data Based on REpresentational State Transfer (REST) Principles. *Comput. Mater. Sci.* **2015**, *97*, 209–215. https://doi.org/10.1016/j.commatsci.2014.10.037.
(61) Tran, R.; Xu, Z.; Radhakrishnan, B.; Winston, D.; Sun, W.; Persson, K. A.; Ong, S. P. Surface Energies of Elemental Crystals. *Sci. Data* **2016**, *3* (1), 160080. https://doi.org/10.1038/sdata.2016.80.

(62) Wolter, H.; Schmidt, M.; Wandelt, K. Surfactant Induced Layer-by-Layer Growth of Cu on Ru(0001) as Revealed by Oscillatory Work Function Changes. *Surf. Sci.* **1993**, *298* (1), 173–186. https://doi.org/10.1016/0039-6028(93)90093-Y.
(63) Azhagurajan, M.; Wen, R.; Lahiri, A.; Kim, Y. G.; Itoh, T.; Itaya, K. Direct Evidence of Homoepitaxial Growth in the Electrodeposition of Au Observed by Ultra-High Resolution Differential Optical Microscopy. *J. Electrochem. Soc.* **2013**, *160* (9), D361–D365. https://doi.org/10.1149/2.057309jes.
(64) Jiang, B.; Xue, H.; Wang, P.; Du, H.; Kang, Y.; Zhao, J.; Wang, S.; Zhou, W.; Bian, Z.; Li, H.; Henzie, J.; Yamauchi, Y. Noble-Metal–Metalloid Alloy Architectures: Mesoporous Amorphous Iridium–Tellurium Alloy for Electrochemical $N_2$ Reduction. *J. Am. Chem. Soc.* **2023**, *145* (11), 6079–6086. https://doi.org/10.1021/jacs.2c10637.
(65) Ernst, H.-J.; Fabre, F.; Lapujoulade, J. Nucleation and Diffusion of Cu Adatoms on Cu(100): A Helium-Atom-Beam Scattering Study. *Phys. Rev. B* **1992**, *46* (3), 1929–1932. https://doi.org/10.1103/PhysRevB.46.1929.
(66) Thiel, P. A.; Evans, J. W. Nucleation, Growth, and Relaxation of Thin Films: Metal(100) Homoepitaxial Systems. *J. Phys. Chem. B* **2000**, *104* (8), 1663–1676. https://doi.org/10.1021/jp9933471.
(67) Sbiaai, K.; Ataalite, H.; Dardouri, M.; Hasnaoui, A.; Fathi, A. Investigation of Growth Mode and Surface Roughness during Homoepitaxial Growth of Silver Metal Using Kinetic Monte Carlo Simulation. *Mater. Today Proc.* **2022**, *66*, 459–465. https://doi.org/10.1016/j.matpr.2022.06.582.
(68) Ronnby, K.; Nolan, M. Predicting the Morphology of Cobalt, Copper, and Ruthenium on TaN for Interconnect Metal Deposition. December 24, 2024. https://doi.org/10.26434/chemrxiv-2024-q9zjf.
(69) Lustemberg, P. G.; Palomino, R. M.; Gutiérrez, R. A.; Grinter, D. C.; Vorokhta, M.; Liu, Z.; Ramírez, P. J.; Matolín, V.; Ganduglia-Pirovano, M. V.; Senanayake, S. D.; Rodriguez, J. A. Direct Conversion of Methane to Methanol on Ni-Ceria Surfaces: Metal–Support Interactions and Water-Enabled Catalytic Conversion by Site Blocking. *J. Am. Chem. Soc.* **2018**, *140* (24), 7681–7687. https://doi.org/10.1021/jacs.8b03809.
(70) Brunet, L.; Fenouillet-Beranger, C.; Batude, P.; Beaurepaire, S.; Ponthenier, F.; Rambal, N.; Mazzocchi, V.; Pin, J.-B.; Acosta-Alba, P.; Kerdiles, S.; Besson, P.; Fontaine, H.; Lardin, T.; Fournel, F.; Larrey, V.; Mazen, F.; Balan, V.; Morales, C.; Guerin, C.; Jousseaume, V.; Federspiel, X.; Ney, D.; Garros, X.; Roman, A.; Scevola, D.; Perreau, P.; Kouemeni-Tchouake, F.; Arnaud, L.; Scibetta, C.; Chevalliez, S.; Aussenac, F.; Aubin, J.; Reboh, S.; Andrieu, F.; Maitrejean, S.; Vinet, M. Breakthroughs in 3D Sequential Technology. In *2018 IEEE International Electron Devices Meeting (IEDM)*; 2018; p 7.2.1-7.2.4. https://doi.org/10.1109/IEDM.2018.8614653.
(71) Kim, J. Y.; Hummel, R. E. Hole and Hillock Formation in Gold Metallizations at Elevated Temperatures Deposited on Titanium, Vanadium, and Other Barrier Layers. *Phys. Status Solidi A* **1990**, *122* (1), 255–273. https://doi.org/10.1002/pssa.2211220125.
(72) Galinski, H.; Ryll, T.; Schlagenhauf, L.; Gauckler, L. J.; Stender, P.; Schmitz, G. Hillock Formation of Pt Thin Films on Single-Crystal Yttria-Stabilized Zirconia. *Phys. Rev. B* **2012**, *85* (12), 125408. https://doi.org/10.1103/PhysRevB.85.125408.
(73) Yang, C.-C.; Li, B.; Baumann, F. H.; Huang, E.; Edelstein, D.; Rosenberg, R. Enhanced Electromigration Resistance through Grain Size Modulation in Copper Interconnects. In *2015 China Semiconductor Technology International Conference*; IEEE: Shanghai, China, 2015; pp 1–5. https://doi.org/10.1109/CSTIC.2015.7153408.

(74) Kasap, S. O. *Principles of Electronic Materials and Devices*, Fourth edition.; McGraw-Hill Education: New York, 2018.

(75) Barmak, K.; Jr, C.; Rodbell, K.; Harper, J. On the Use of Alloying Elements for Cu Interconnect Applications. *J. Vac. Sci. Technol. B Microelectron. Nanometer Struct.* **2006**, *24*, 2485–2498. https://doi.org/10.1116/1.2357744.

(76) Dutta, S.; Sankaran, K.; Moors, K.; Pourtois, G.; Van Elshocht, S.; Bömmels, J.; Vandervorst, W.; Tőkei, Z.; Adelmann, C. Thickness Dependence of the Resistivity of Platinum-Group Metal Thin Films. *J. Appl. Phys.* **2017**, *122* (2), 025107. https://doi.org/10.1063/1.4992089.

(77) Gall, D. Metals for Low-Resistivity Interconnects. In *2018 IEEE International Interconnect Technology Conference (IITC)*; 2018; pp 157–159. https://doi.org/10.1109/IITC.2018.8456810.

(78) Sung, J. Y.; Jo, G.-Y.; Moon, S. M.; Lee, C. H.; Lim, Y.; Yun, Y.; Lee, S. H.; Lee, J.-Y.; Yang, D.-J.; Kim, S. W.; Lee, S. W. Demonstration of NiCo as an Alternative Metal for Post-Cu Interconnects. *ACS Nano* **2025**, *19* (7), 7253–7262. https://doi.org/10.1021/acsnano.4c17216.

(79) Chen, L.; Ando, D.; Sutou, Y.; Gall, D.; Koike, J. NiAl as a Potential Material for Liner- and Barrier-Free Interconnect in Ultrasmall Technology Node. *Appl. Phys. Lett.* **2018**, *113* (18), 183503. https://doi.org/10.1063/1.5049620.

For Table of Contents Only

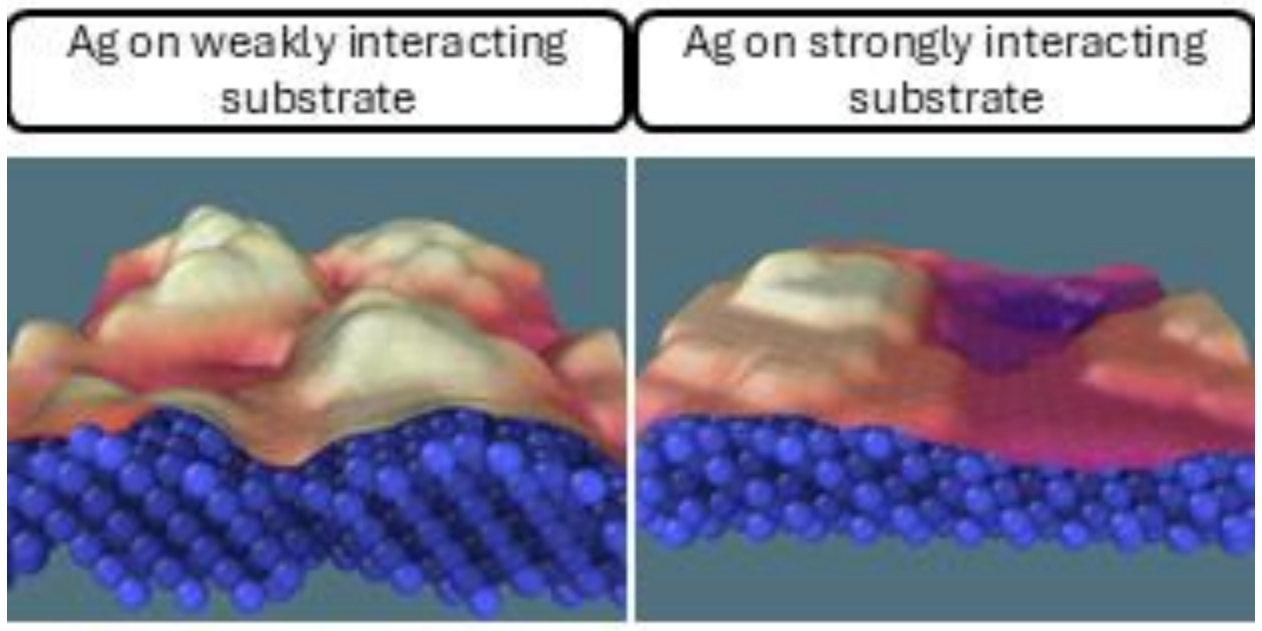